%% file: FullBC_v3.tex
\newif\ifconfver
 \confvertrue        

\newif\ifplainver  
 \plainverfalse

\ifplainver
    \confverfalse   
\fi

\ifconfver
     \documentclass[10pt,twocolumn,twoside]{IEEEtran}
\else
    \ifplainver
        \documentclass[11pt]{article}
        \usepackage{fullpage}
    \else
        \documentclass[11pt,draftcls,onecolumn]{IEEEtran}
    \fi
\fi
\usepackage{calc,amsfonts,amssymb,amsmath ,bm,url,theorem,graphicx,cite,shortcuts_OPT,enumitem,booktabs}

\usepackage{colortbl}

\usepackage{algorithm}
\usepackage{algorithmic}
\usepackage{psfrag,subcaption,float,hyperref,bbm,cleveref}
\hypersetup{
  colorlinks   = true, 
  urlcolor     = blue, 
  linkcolor    = blue, 
  citecolor    = blue   
}

\usepackage{tikz}
\usetikzlibrary{
  arrows.meta,
  chains,
  matrix,
  positioning,shapes.geometric}

\newtheorem{Fact}{Fact}
\newtheorem{Lemma}{Lemma}

\newtheorem{Corollary}{Corollary}

\newtheorem{assumption}{Assumption}
\newtheorem{Remark}{Remark}

\theorembodyfont{\rmfamily}
\newtheorem{Exa}{Example}

\usepackage{pgfplots}
\usetikzlibrary{arrows,shapes,calc,tikzmark,backgrounds,matrix,decorations.markings,backgrounds}
\usepgfplotslibrary{fillbetween}
 
\pgfplotsset{compat=1.3}

\usepackage{relsize}
\tikzset{fontscale/.style = {font=\relsize{#1}}
    }

\definecolor{lavander}{cmyk}{0,0.48,0,0}
\definecolor{violet}{cmyk}{0.79,0.88,0,0}
\definecolor{burntorange}{cmyk}{0,0.52,1,0}

\definecolor{asuorange}{rgb}{1,0.699,0.0625}
\definecolor{asured}{rgb}{0.598,0,0.199}
\definecolor{asuborder}{rgb}{0.953,0.484,0}
\definecolor{asugrey}{rgb}{0.309,0.332,0.340}
\definecolor{asublue}{rgb}{0,0.555,0.836}
\definecolor{asugold}{rgb}{1,0.777,0.008}

\tikzstyle{latent}=[draw,circle, black!70, fill = black!30,
                       text=violet,minimum width=10pt]
\tikzstyle{superpeers}=[draw,circle, asublue!80!white, fill = asublue!50!white,
                       text=violet,minimum width=10pt]
\tikzstyle{susceptible}=[draw,circle, red, fill = red!50,
                       text=violet,minimum width=10pt]

\definecolor{color0}{rgb}{0,0.75,0.75}
\newenvironment{customlegend}[1][]{%
  \begingroup
  \csname pgfplots@init@cleared@structures\endcsname
  \pgfplotsset{#1}%
}{%
  \csname pgfplots@createlegend\endcsname
  \endgroup
}%
\def\addlegendimage{\csname pgfplots@addlegendimage\endcsname}

    \makeatletter
    \def\multilimits@{\bgroup
  \Let@
  \restore@math@cr
  \default@tag
 \baselineskip\fontdimen10 \scriptfont\tw@
 \advance\baselineskip\fontdimen12 \scriptfont\tw@
 \lineskip\thr@@\fontdimen8 \scriptfont\thr@@
 \lineskiplimit\lineskip
 \vbox\bgroup\ialign\bgroup\hfil$\m@th\scriptstyle{##}$\hfil\crcr}
    \def\Sb{_\multilimits@}
    \def\endSb{\crcr\egroup\egroup\egroup}
\makeatother

\usepackage{pgfplots}   
\usepgfplotslibrary{units}
\usetikzlibrary{spy,backgrounds}

\makeatletter
\DeclareRobustCommand*\cal{\@fontswitch\relax\mathcal}
\makeatother

\begin{document}
\title{Detecting Central Nodes from Low-rank Excited Graph Signals via Structured Factor Analysis}
\author{Yiran He, Hoi-To Wai\thanks{A preliminary version of this work has been presented at ICASSP 2020, Barcelona, Spain \cite{Wai2020BCest}. 
Y.~He and  H.-T.~Wai are with the Department of SEEM, The Chinese University of Hong Kong, Shatin, Hong Kong SAR of China. E-mails: \url{yrhe@se.cuhk.edu.hk}, \url{htwai@se.cuhk.edu.hk}.}} 

\maketitle
\begin{abstract}
This paper treats a blind detection problem to identify the central nodes in a graph from filtered graph signals. Unlike prior works which impose strong restrictions on the data model, we only require the underlying graph filter to satisfy a low pass property with a generic low-rank excitation model. We treat two cases depending on the low pass graph filter's strength. When the graph filter is strong low pass, i.e., it has a frequency response that drops sharply at the high frequencies,  we show that the principal component analysis (PCA) method detects central nodes with high accuracy. For general low pass graph filter, we show that the graph signals can be described by a structured factor model featuring the product between a low-rank plus sparse factor and an unstructured factor. We propose a two-stage decomposition algorithm to learn the structured factor model via a judicious combination of the non-negative matrix factorization and robust PCA algorithms. We analyze the identifiability conditions for the model which lead to accurate central nodes detection. Numerical experiments on synthetic and real data are provided to support our findings. We demonstrate significant performance gains over prior works.\end{abstract}

\begin{IEEEkeywords} 
graph signal processing, low pass graph filter, centrality estimation, nonnegative matrix factorization
\end{IEEEkeywords}

\section{Introduction}
Recently, the study of network science has proliferated in research fields such as social science, biology, and data science, where it has offered new insights through studying network data from a graph's perspective. A popular subject is to identify central nodes through the \emph{node centrality} \cite{newman2018networks}. Node centrality ranks the relative importance of nodes with respect to the graph topology, and it is useful for identifying influential people in a social network, frequently visited webpages on the Internet, stocks that are important driving forces in the market, etc. \cite{borgatti2005centrality}. 
A number of centrality measures such as degree centrality, betweenness centrality, eigen-centrality, etc., have been studied \cite{bloch2019centrality}. Among others, the eigen-centrality measure, defined as the top eigenvector of the graph adjacency matrix, is popular as it takes into account the importance of the node's neighbors as well as the node's own degree. 
These appealing features have led to the wide adoption of the eigen-centrality measure, for instance, the PageRank model \cite{DBLP2014}.

Evaluating the eigen-centrality typically requires knowledge of the graph topology. In this paper, we focus on using a data-driven approach to learn the eigen-centrality. Particularly, we develop algorithms that use just \emph{graph signals} observed on the nodes, e.g., opinions on social networks, or return records in stock market, etc., to detect nodes with high eigen-centrality. We call this task the \emph{blind central nodes detection problem}.
The blind detection problem is related to the fast growing literature in graph learning \cite{dong2019learning, mateos2019connecting}. In particular, a natural heuristic is to first perform graph learning, and then rank the eigen-centrality using the estimated graph. 
Such a heuristic is prone to errors since the graph learning algorithms typically require strong assumptions to provide reliable estimates of the graph. 
For example, they assume precise knowledge of generative models such as graphical Markov random field \cite{friedman2008sparse}, non-linear dynamical systems \cite{ioannidis2019semi,wai2019joint}, etc.. As an alternative, the notion of \emph{filtered graph signals} which is rooted in graph signal processing (GSP) \cite{Sand2013DSP, ortega2018graph} provides a flexible model through viewing graph signals as generated by filtering the excitation through an unknown graph filter. Based on this model, graph learning algorithms have been proposed which exploit smoothness \cite{dong2016learning}, spectral template \cite{segarra2017network}, graph structure priors \cite{egilmez2017graph, Yuan2019BCD}, etc.. 
However, further restrictions on the excitation are needed for the above algorithms. For instance, it is assumed in \cite{dong2016learning, segarra2017network, egilmez2017graph, Yuan2019BCD} that the excitation is white noise with independent input at each node. 
The latter assumption which implies full-rank observations may not hold in practice \cite{udell2019big}.

The approach pursued by the current paper is to detect central nodes \emph{directly} from filtered graph signals, thereby side-stepping the error-prone graph learning stage. 
Doing so allows us to relax some restrictions required by previous works on graph learning. Notably, our approach applies to general graph signal models with possibly low-rank excitation. Our key assumption is that the observed graph signals are filtered by a \emph{low pass graph filter} that has a frequency response which drops at the high frequencies, {\blue while the excitation to the graph filter is independent of the graph}. This property is commonly found in graph data from applications such as economics, social networks, and power systems, etc.~\cite{ramakrishna2020user}.
Depending on the strengths of the low pass graph filter, we treat two cases separately. With a \emph{strong} low pass graph filter whose frequency response drops \emph{sharply} at the high frequencies [cf.~$\eta \approx 0$ in \Cref{assump:lowpass}], we show that the principal component analysis (PCA) method detects the central nodes accurately. Note that this case corresponds to graph signals that are \emph{smooth} with respect to the graph topology. With a \emph{general} low pass graph filter which may or may not be strong low pass [cf.~$\eta < 1$ in \Cref{assump:lowpass}], the blind detection problem becomes more challenging to tackle. As a trade-off, we are motivated by applications in economics and social networks to concentrate on an additional assumption that the excitation graph signals lie in a subspace with the basis given by a sparse and non-negative matrix. Subsequently, the filtered graph signals are treated by a novel \emph{structured factor analysis} model featuring the product between a `low-rank plus sparse' factor and an unstructured factor.  
We develop a two-stage decomposition algorithm leveraging a sparse non-negative matrix factorization (NMF) criterion and the robust PCA (RPCA) method. We show that the central nodes of the underlying graph can be detected accurately through analyzing the identifiability of the factor model and a boosting property of the low pass graph filter. To our best knowledge, this is the first application of the NMF and RPCA to detect central nodes from graph signals. 

Notice that NMF and RPCA have been separately applied in other domains such as blind source separation, topic modeling, and video surveillance  \cite{fu2019nonnegative,agarwal2012noisy,candes2011robust}. The theoretical guarantees therein also have different focuses from the current paper.
We remark that the approach of side-stepping graph learning is inspired by \cite{Schaub2020BCD,wai2019blind,Hoffmann2020BCD,roddenberry2020exact,xing2020community} on the \emph{blind community detection} problem. In comparison, this paper is the first to handle blind central nodes detection under the challenging case with unknown excitation. Lastly, blind centrality ranking has been treated in \cite{roddenberry2020blind,roddenberry2020blinda} using the PCA method. Assuming white noise excitation, these works provided the sample complexity analysis on ranking the eigen-centrality of nodes. In comparison, our work assumes low-rank excitation that yields a strictly more general model.
 
\noindent \textbf{Contributions.} This paper treats the blind detection of central nodes using low-rank excited low pass filtered graph signals. Our contributions are as follows:\vspace{-.2cm}
\begin{itemize}[leftmargin=*, itemsep=.1cm]
    \item We show that the PCA method accurately identifies the central nodes under the conditions that the graph signals are generated by a strong low pass graph filter [cf.~$\eta \approx 0$ in \Cref{assump:lowpass}] or the excitation is white noise. Our analysis demonstrates an explicit dependence of the detection performance on factors such as the sample size.
    \item We recognize a structured factor model for the filtered graph signals via an intrinsic decomposition featuring a low-rank plus sparse factor and an unstructured factor under the condition that the graph signals are generated by a general low pass graph filter [cf.~$\eta < 1$ in \Cref{assump:lowpass}] which may or may not be strong low pass. Note we have concentrated on a special case where the excitation graph signals lie in a subspace with sparse and non-negative basis, as motivated by applications in economics and social networks.
    \item We provide the first identifiability analysis on the structured factor model involving filtered graph signals with unknown excitation. We show that the identification performance depends on the low pass ratio of the graph filter and the ratio of the excitation's rank to the graph's size. 
    Inspired by the identifiability analysis, we propose to combine a sparse NMF method with the RPCA method to yield a two-stage algorithm for efficient central nodes detection.
    \item We carry out numerical experiments on real and synthetic data to verify our theoretical results. Compared to existing works, our methods can consistently detect the central nodes with a lower error rate on synthetic data and {\blue higher correlation with the global behavior on real data}.
\end{itemize}
Compared to the conference version \cite{Wai2020BCest}, this paper includes the case with general low pass filter and unknown excitation. We also included an extended set of numerical experiments. 

\noindent \textbf{Organization.} The rest of this paper is organized as follows. In Section~\ref{sec:model}, we introduce the graph signal model. {Two application examples are discussed and their graph signal models are derived.} In Section~\ref{sec:struct}, {we consider strong low pass graph filter and prove that the PCA method can detect central nodes with high accuracy. In Section~\ref{sec:weak}, we consider general low pass graph filter and propose a two-stage algorithm motivated by identifiability conditions with NMF specialized to leverage the graph signal structure. Finally, we discuss the practical implementation issues in Section~\ref{sec:practical}, and present numerical experiments to support our findings in Section~\ref{sec:exp}. }

\noindent \textbf{Notations.} We use boldfaced character (resp.~boldfaced capital letter) to denote vector (resp.~matrix). For a matrix ${\bm X} \in \Re^{m \times n}$, we use $[{\bm X}]_{i,j}$ to denote its $(i,j)$th entry. For any vector ${\bm x} \in \Re^n$, $\| {\bm x} \|$, $\| {\bm x} \|_1$ denote the Euclidean, $\ell_1$ norm, respectively; for any matrix ${\bm X} \in \Re^{m \times n}$, $\| {\bm X} \|$ denotes the spectral norm, $\| {\bm X} \|_\star$ denotes the nuclear norm, and $\|{\bm X}\|_F = \sqrt{\sum_{i,j} X_{ij}^2}$ denotes the Frobenius norm. For a random vector ${\bm y}_\ell$, its covariance is denoted as the matrix ${\rm Cov}({\bm y}_\ell) = \EE[ ( {\bm y}_\ell - \EE[{\bm y}_\ell] ) ( {\bm y}_\ell - \EE[{\bm y}_\ell] )^\top ]$.\vspace{-.2cm}

\begin{figure}[t]
\centering
\includegraphics[width=.45\linewidth]{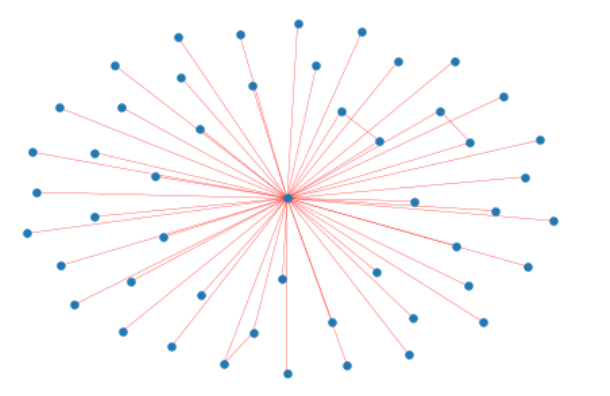}~\includegraphics[width=.45\linewidth]{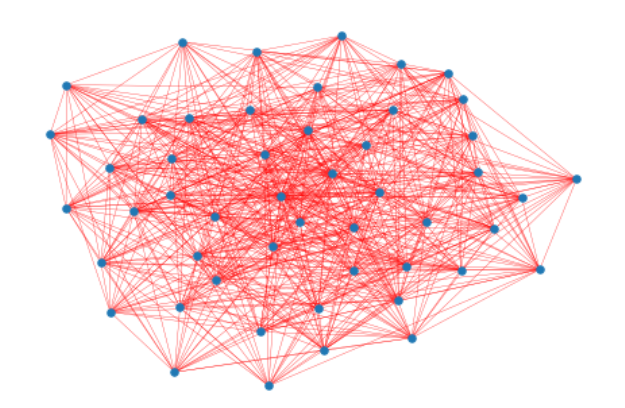}
\caption{(Left) Graph with a central hub node: ${\lambda_2} / {\lambda_1}\approx 0.1238$. (Right) an Erdos-Renyi graph that does not have central node(s) with significantly higher degree: ${\lambda_2} / {\lambda_1} \approx 0.9189$.}\vspace{-.4cm}
\label{fig:lothighcon}
\end{figure}

\section{Graph Signal Model}\label{sec:model}\vspace{-.2cm}
Consider an undirected, connected graph $G = ({\cal V},{\cal E}, {\bm A})$ with $n$ nodes given by ${\cal V} := \{1, \dots, n\}$ and edge set ${\cal E} \subseteq {\cal V} \times {\cal V}$. The matrix ${\bm A} \in \Re_+^{n \times n}$ is a weighted adjacency matrix such that $ A_{ij} = A_{ji} > 0$ if and only if $(i,j) \in E $; otherwise, $A_{ij} = 0$. The symmetric matrix ${\bm A}$ admits an eigenvalue decomposition as ${\bm A} = {\bm V} \bm{\Lambda} {\bm V}^\top$, where ${\bm V}$ is an orthogonal matrix and $\bm{\Lambda} = {\rm Diag}( \bm{\lambda} )$ is a diagonal matrix of the eigenvalues. For simplicity, we assume the eigenvalues are ordered as $\lambda_1 > \lambda_2 \geq \cdots \geq \lambda_n$. 
{To specify the scenario of interest, we focus on graphs with $C$ central nodes of high inter/intra-connectivity, for some $C \ll n$. As studied by \cite{cucuringu2016detection}, such graphs can be characterized by large spectral gap with $\lambda_1 \gg \lambda_2$ [cf.~Fig.~\ref{fig:lothighcon}].}
Our aim is to detect these central nodes defined via the eigen-centrality vector \cite{newman2018networks}:\vspace{-.05cm}
\begin{equation} \label{eq:eigcen}
{\bm c}_{\sf eig} := {\sf TopEV}({\bm A}) = {\bm v}_1,\vspace{-.1cm}
\end{equation}
where ${\sf TopEV}({\bm A})$ denotes the top eigenvector of ${\bm A}$. 
{Accordingly, we denote the \emph{top-$C$ central nodes} as the set of $C$ nodes whose magnitudes $|[{\bm c}_{\sf eig}]_i|$ are the highest.}

{We observe a set of vectors $\{ {\bm y}_\ell \}_{\ell=1}^m$ generated from a certain process on $G$ treated as \emph{filtered graph signals} \cite{Sand2013DSP}. 
The graph signal defined on $G$ is an $n$-dimensional vector ${\bm y} \in \RR^n$ such that $y_i$ denotes the signal on node $i$.} On the other hand, a graph filter is described as a polynomial of the graph shift operator (GSO):
\begin{equation} \label{eq:graphfil}
{\cal H}({\bm A}) = \sum_{t=0}^{T-1} h_t {\bm A}^t = {\bm V} \left(\sum_{t=0}^{T-1}h_t \bm{\Lambda}^t \right) {\bm V}^\top \in \Re^{n\times n},
\end{equation}
where $\{ h_t \}_{t=0}^{T-1}$ are the filter coefficients and {\blue $T \in \ZZ_+ \cup \{ \infty\}$} is the filter's order. In this paper, we took the adjacency matrix ${\bm A}$ as the GSO. Accordingly, the graph Fourier transform (GFT) of ${\bm y}$ is given by $\tilde{\bm y} = {\bm V}^\top {\bm y}$ \cite{sandryhaila2014discrete}. With this choice of GFT basis, we define the diagonal matrix $h(\bm{\Lambda}) = \sum_{t=0}^{T-1}h_t \bm{\Lambda}^t$, whose diagonal entries are  the \emph{frequency responses} of the graph filter. The $i$th diagonal $h(\lambda_i) \eqdef [h(\bm{\Lambda})]_{i,i}$ denotes the frequency response corresponding to the $i$th graph frequency.  

Finally, the observations $\{ {\bm y}_\ell \}_{\ell=1}^m$ are modeled as the graph filter's outputs subject to the excitation $\{ {\bm x}_\ell \}_{\ell=1}^m$:
\beq \label{eq:y=Hx}
{\bm y}_\ell = {\cal H}( {\bm A} ) {\bm x}_\ell + {\bm w}_\ell,~\ell=1,...,m,
\eeq
where ${\bm w}_\ell \sim {\cal N}( {\bm 0}, \sigma^2 {\bm I} )$ represents the modeling error and measurement noise. We assume that ${\bm w}_\ell, {\bm w}_{\ell'}$ are independent if $\ell \neq \ell'$. 
{\blue The excitation $\{ {\bm x}_\ell \}_{\ell=1}^m$ represent the external stimuli inflicted on $G$ and they are assumed to be independent of the graph, see \eqref{eq:x=Bz} for further discussion.}
The main task of this paper is to tackle the \emph{blind central nodes detection} problem, where we identify the top-$C$ central nodes using just $\{ {\bm y}_\ell \}_{\ell=1}^m$.

To identify the central nodes from $\{ {\bm y}_\ell \}_{\ell=1}^m$, our key assumption is that the graph filter ${\cal H}({\bm A})$ satisfies a \emph{low pass} property \cite{ramakrishna2020user}, i.e., its frequency response drops at the high graph frequencies. Particularly, we assume:\vspace{-.1cm}
\begin{assumption} \label{assump:lowpass}
The graph filter ${\cal H}({\bm A})$ is $1$-low pass such that its frequency response satisfies:
\beq \label{eq:lowpass}
\eta := {\max_{j=2,...,n} |h(\lambda_j)|} / {|h(\lambda_1)|} < 1,
\eeq 
where $\eta$ is called the low pass ratio {\blue that quantifies the stopband attenuation, see Fig.~\ref{fig:stopband}}. \vspace{-.1cm}
\end{assumption}
Observe that condition \eqref{eq:lowpass} {\blue describes a (1-)low pass graph filter whose cutoff frequency occurs at $\lambda_1$, i.e., the lowest graph frequency.} It is imposed on the frequency response function $h(\lambda)$ which requires $| h(\lambda) |$ to be strictly maximized at the lowest graph frequency $\lambda_1$ over $\{\lambda_1,...,\lambda_n\}$, justifying its low pass interpretation. {\blue For convenience, we will refer to such filter as simply low pass graph filter in the rest of this paper.}
The low pass assumption is common for modeling practical network data; see the overview in \cite{ramakrishna2020user}. 
Examples of low pass graph filter include the $\alpha$-diffusion filter that is inspired by heat diffusion \cite{thanou2017learning}, given as ${\cal H}( {\bm A} ) = e^{\alpha {\bm A}}$ with $\alpha > 0$; the infinite impulse response (IIR) filter given by ${\cal H}( {\bm A}) = ({\bm I} - c {\bm A} )^{-1}$ with $c>0$.

{\blue The low pass ratio $\eta < 1$ characterizes the amount of stopband attenuation of the low pass graph filter, which we shall use to describe the \emph{strength} of the low pass graph filter.} As a convention, we say that the low pass filter is \emph{strong} when $\eta \approx 0$; otherwise, the low pass filter is \emph{weak} when $\eta \approx 1$. 
For example, the diffusion filter ${\cal H}( {\bm A} ) = e^{\alpha {\bm A}}$ is strong low pass for large $\alpha$, particularly, the low pass ratio decreases to zero exponentially with $\alpha$; the IIR filter ${\cal H}( {\bm A}) = ({\bm I} - c {\bm A} )^{-1}$ is weak low-pass for small $c$.
\vspace{-.2cm}

{\blue \begin{figure}[t]
\centering
{\sf 
\resizebox{!}{.5\linewidth}{\input{stopband}}
}\vspace{-0.4cm}
\caption{ Frequency response $h(\lambda_i)$ against eigenvalue $\lambda_i$ under ({\color{green!50!black}Green}) weak low pass filter ${\cal H}_{w}({\bm A})=({\bm I}-0.01{\bm A})^{-1}$ and ({\color{blue}Blue}) strong low pass filter ${\cal H}_{s}({\bm A})=\frac{1}{7}e^{0.2{\bm A}}$ with a core–periphery structure $\lambda_2/\lambda_1\approx0.12$. The red shadow area indicates the transition region with \textbf{cutoff frequency} $\lambda_1$ while the black shadow area indicates the stopband region. The low pass ratios are $\eta_{\sf w} \approx 0.92$ and $\eta_{\sf s} \approx 0.2$. The lengths of the black arrows show the stopband attenuation. \emph{Notice that the $x$-axis is flipped such that $\lambda_1>\lambda_2\geq\dots\geq\lambda_n$}.}\label{fig:stopband}
\end{figure}
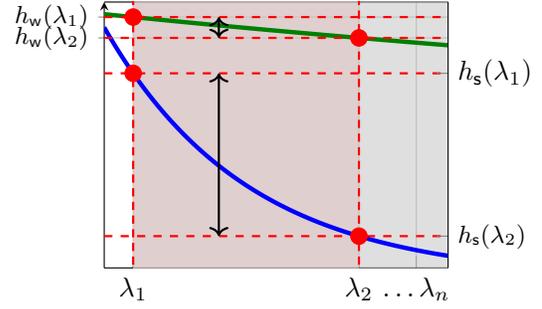}
 
\subsection{The Excitation Graph Signals and Applications}\vspace{-.2cm} \label{sec:case}
{As discussed in the Introduction, a majority of prior works on graph learning have imposed various restrictions on the excitation graph signals $\{ {\bm x}_\ell \}_{\ell=1}^m$. For example, in \cite{segarra2017network, roddenberry2020blind, roddenberry2020blinda}, it is assumed that the excitation are zero-mean white noise with $\EE[ {\bm x}_\ell {\bm x}_\ell^\top ] = {\bm I}$, i.e., the excitation at each node are independent. 
In practice, there can be correlation between the excitation at different nodes in $G$. 

To study a general excitation model, we assume without loss of generality (w.l.o.g.) that ${\bm x}_\ell$ lies in a $k$-dimensional subspace spanned by $\{ {\bm b}_j \}_{j=1}^k$. Here, $1 \leq k \leq n$ is the \emph{excitation's rank}. We write:
\begin{equation}\label{eq:x=Bz} \textstyle
 {\bm x}_\ell = \sum_{j=1}^k {\bm b}_j z_{\ell,j} =  {\bm B} {\bm z}_\ell,~~\ell=1,...,m,
\end{equation}
such that ${\bm z}_\ell \in \RR^k$ represents the latent parameters of the excitation. The columns of the matrix ${\bm B}$ form the basis for the subspace that $\{ {\bm x}_\ell \}_{\ell=1}^m$ lie in.
When $k =n, {\bm B} = {\bm I}$, $\EE[ {\bm z}_\ell ( {\bm z}_\ell )^\top ] = {\bm I}$, we recover the white noise model found in \cite{segarra2017network, roddenberry2020blind, roddenberry2020blinda}. 
In general, the $j$th column vector ${\bm b}_j$ of ${\bm B} \in \RR^{n \times k}$ represents the influence profile from an external source $z_{\ell,j}$ to impact on the graph. The excitation ${\bm x}_\ell$ is a superposition of these influences from external sources. {\blue Notice that in general, ${\bm B}$ does not depend on $G$ as the excitation models external stimuli, see the examples below.}

As we shall demonstrate in the forthcoming sections, the performance of blind central nodes detection is affected by ${\bm B} \neq {\bm I}$, especially if the graph filter is weak low pass. This motivates us to further investigate the \emph{structure} of ${\bm B}$ in common applications. Specifically, we show that popular economics and social network models can be approximated as filtered graph signals obeying \eqref{eq:y=Hx}, \eqref{eq:x=Bz}. The basis ${\bm B}$ in these cases can be modeled as a {sparse and non-negative matrix}\footnote{We remark that the typical sparse matrix ${\bm B}$ mentioned in our discussions have around 10\%--20\% of non-zero entries.}.
}

\begin{Exa}\emph{(Opinions)} \label{ex:op}
We consider a social network denoted by $G = ({\cal V} , {\cal E}, {\bm A} )$, where $A_{ij} = A_{ji} \geq 0$ denotes the trust strength between agents $i,j$. As shown in \cite[Example 3]{wai2019blind}, with the DeGroot model \cite{degroot1974reaching}, the steady-state opinions on the $\ell$th topic, ${\bm y}_\ell$, can be shaped by $k$ stubborn agents ($k \ll n$), denoted as
\beq \label{eq:social}
{\bm y}_\ell = ({\bm I} - {\bm A})^{-1} {\bm B} {\bm z}_\ell,
\eeq
where ${\bm z}_\ell \in \RR_+^k$ are the opinions held by the stubborn agents and ${\bm B} \in \RR_+^{n \times k}$ describes the mutual trusts between stubborn and non-stubborn agents. The opinions, which model the probabilities of an action, are non-negative and the stubborn agents are those who influence the non-stubborn agents without being influenced. From \eqref{eq:social}, it is easy to observe that the steady-state opinion data follows a special case of \eqref{eq:y=Hx}, \eqref{eq:x=Bz}. {\blue For sparse social networks with a small number of stubborn agents}, the mutual trust matrix ${\bm B}$ is non-negative, tall and sparse.\hfill $\square$
\end{Exa}

\begin{Exa} \label{ex:stock}
\emph{(Stock Returns)}
We consider an inter-stock influence network described by $G = ({\cal V} , {\cal E}, {\bm A} )$, whose node set consists of $n$ stocks and $(i,j) \in {\cal E}$ indicates that stocks $i$, $j$ are dependent on each other, e.g., when there are business ties. {The graph is endowed with a weighted adjacency matrix ${\bm A}$. A popular model from \cite{billio2012econometric} suggests that on day $\ell$, the time series of stock prices, $\{ {\bm s}_{\ell,t} \}_t$, may evolve as 
\beq \label{eq:granger}
{\bm s}_{\ell, t+1} = {\bm x}_\ell + {\bm A} {\bm s}_{\ell, t} + {\bm e}_{\ell, t},
\eeq
where ${\bm e}_{\ell,t}$ is a zero-mean noise and ${\bm x}_\ell$ is the mean vector\footnote{The model is slightly modified from \cite[Eq.~(9)]{billio2012econometric} as we consider unnormalized data with non-zero mean.} for stock prices on day $\ell$. Notice that \eqref{eq:granger} describes the evolution of stock prices at a fast timescale. As such, the daily returns of the $n$ stocks ${\bm y}_\ell$, defined as the ratio between closing and opening prices in a day, are given by ${\bm s}_{\ell,T}$ with $T \gg 1$. We observe:
\beq \label{eq:filter_stock}
\begin{split}
{\bm y}_\ell & \textstyle = {\bm s}_{\ell,T} = {\bm A}^T {\bm s}_{\ell,0} + \sum_{t=0}^T {\bm A}^t \{ {\bm x}_\ell + {\bm e}_{\ell,t} \} \\
& \approx ({\bm I} - {\bm A})^{-1} {\bm x}_\ell,
\end{split}
\eeq
where the approximation assumes that $\| {\bm A} \| < 1$, ${\bm e}_{\ell,t} \approx 0$, and the opening stock prices ${\bm s}_{\ell,0}$ are normalized to ${\bf 1}$.
Thus, the model \eqref{eq:filter_stock} is a special case of \eqref{eq:y=Hx}. The mean vector ${\bm x}_\ell$ may describe the state of the businesses on day $\ell$. 

As shown in \cite{brown1989number}, the stock returns are dominated by a few factors implying a low-rank data model. This motivates us to model ${\bm x}_\ell$ as in \eqref{eq:x=Bz} with $k \ll n$, where ${\bm B} \in \RR^{n \times k}$. 
Additionally, the latent parameters ${\bm z}_\ell \in \RR^k$ encode the state-of-the-world on day $\ell$. An example is that for some $j \in \{1,..., k\}$, $z_{\ell,j} \geq 0$ is the attention level of the market to `oil crisis', then the stocks in the energy sector will be directly affected by $z_{\ell,j}$. Moreover, ${\bm b}_j$ is a vector supported only on these stocks. Such sector-specific influence structure suggests us to consider ${\bm B}$ as a sparse matrix. {\blue Notice that it is reasonable to further assume that ${\bm B} \geq {\bm 0}$ through considering a constant shift to the excitation.}} \hfill $\square$
\end{Exa}\vspace{-.1cm}

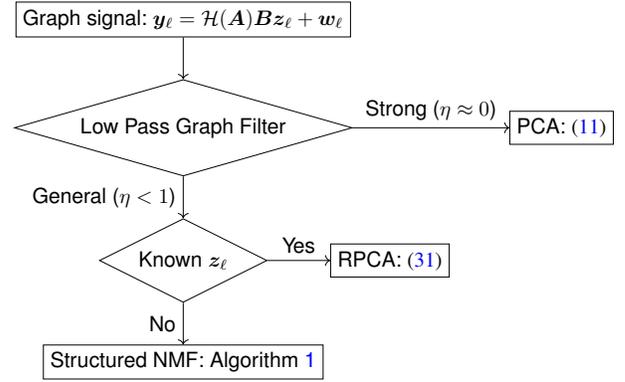
\begin{figure}[t]
\begin{center}\resizebox{.9\linewidth}{!}{\sf
\begin{tikzpicture}[node distance=10pt]
  \node[draw] (start)   {Graph signal: ${\bm y}_\ell = {\cal H}({\bm A}) {\bm B} {\bm z}_\ell + {\bm w}_\ell$};
  \node[draw, diamond, aspect=3.5,below=20pt of start] (lowpass)  {Low Pass Graph Filter};
  \node[draw, right=75pt of lowpass] (yeslp)  {PCA: \eqref{eq:pca}};
  \node[draw,  diamond, aspect=2,below=20pt of lowpass] (zknown)  {Known ${\bm z}_\ell$};
  \node[draw, right=30pt of zknown] (yesz)  {RPCA: \eqref{eq:rpca}};
  \node[draw, below=20pt of zknown] (end) {Structured NMF: Algorithm~\ref{alg:mega}};
  \draw[->] (start)  -- (lowpass);
  \draw[->] (lowpass) -- node[above] {Strong ($\eta \approx 0$)} (yeslp);
  \draw[->] (lowpass) -- node[left] {General ($\eta < 1$)} (zknown);
  \draw[->] (zknown) -- node[above]  {Yes} (yesz);
  \draw[->] (zknown) -- node[left] {No} (end);
\end{tikzpicture}}\vspace{-.2cm}
\end{center}
\caption{Summary of the proposed blind central nodes detection framework. The above shows the studied algorithms and their suitable applications with respect to the low pass graph filter's strength and data availability. We discuss how to distinguish between strong/general low pass graph filter in \Cref{sec:practical}.}\label{fig:overview}\vspace{-.2cm}
\end{figure}

In the forthcoming sections, we discuss the blind central nodes detection problems when the underlying low pass graph filter is of different strengths. An overview of our strategies for different scenarios is presented in Fig.~\ref{fig:overview}.

\vspace{-.2cm}
\section{Detection with Strong Low Pass Filter}
\vspace{-.2cm}
\label{sec:struct}
This section discusses the blind central nodes detection problem under a strong low pass graph filter, i.e., when $\eta \approx 0$ in Assumption~\ref{assump:lowpass} such that the frequency response decreases sharply over the transition region $[\lambda_2, \lambda_1]$, see Fig.~\ref{fig:stopband}. Importantly, our analysis also highlights the role of the low pass ratio $\eta$ in estimating the eigen-centrality vector.

To fix ideas, we concentrate on the covariance matrix of the filtered graph signals and explain the insight behind our central nodes detection method. W.l.o.g., in this section we assume that ${\rm Cov}({\bm z}_\ell) = {\bm I}$ and observe that
\begin{equation} \label{eq:y_cov}
\begin{split}
    {\rm Cov}({\bm y}_\ell) 
    & = {\bm V} h( \bm{\Lambda} ) {\bm V}^\top {\bm B} {\bm B}^\top {\bm V} h( \bm{\Lambda} ) {\bm V}^\top + \sigma^2 {\bm I}.
\end{split}
\end{equation}
{The top eigenvector of ${\rm Cov}({\bm y}_\ell)$ is different from ${\bm c}_{\sf eig}$ in general if ${\bm B} \neq {\bm I}$. A special case is observed}
when $\eta \approx 0$, where $h( \bm{\Lambda} ) = \sum_{t=0}^{T-1} h_t \bm{\Lambda}^t$ can be approximated by the rank-one matrix $h(\lambda_1) {\bm e}_1 {\bm e}_1^\top$. Recalling that ${\bm v}_1 = {\bm c}_{\sf eig}$, we have
\beq \label{eq:lowpass_approx} 
{\rm Cov}({\bm y}_\ell) 
\approx \|{\bm B}^\top {\bm c}_{\sf eig}\|^2 |h(\lambda_1)|^2 \, {\bm c}_{\sf eig} {\bm c}_{\sf eig}^\top + \sigma^2 {\bm I}.
\eeq
{As a consequence, the top eigenvector of ${\rm Cov}({\bm y}_\ell)$ shall be close to ${\bm c}_{\sf eig}$ when the noise is small. The above calculations motivated us to detect central nodes from $\{ {\bm y}_\ell \}_{\ell=1}^m$ by taking the top eigenvector of sampled covariance, as follows.}

{\noindent \textbf{PCA Method.} The principal component analysis (PCA) method proceeds by calculating the vector:}
\beq \label{eq:pca} \textstyle
\widehat{\bm v}_1 = {\sf TopEV} \big( \widehat{\bm C}_y \big) ~~\text{where}~~ \widehat{\bm C}_y \eqdef \frac{1}{m} \sum_{\ell=1}^m {\bm y}_\ell {\bm y}_\ell^\top
\eeq
{as a surrogate to ${\bm c}_{\sf eig}$. Subsequently, we select $C$ nodes with the highest-$C$ magnitudes from the vector $\widehat{\bm v}_1$ as the detection output of central nodes.}

{To analyze the performance of the PCA method, we shall bound the error in estimating the eigen-centrality vector by $\widehat{\bm v}_1$.} We first let $\widehat{\bm v}_1^\top {\bm c}_{\sf eig} \geq 0$ w.l.o.g.. We set the noiseless covariance matrix as $\overline{\bm C}_y := {\bm V} h( \bm{\Lambda} ) {\bm V}^\top {\bm B} {\bm B}^\top {\bm V} h( \bm{\Lambda} ) {\bm V}^\top$ and the finite sample error as $\bm{\Delta} := \widehat{\bm C}_y - \overline{\bm C}_y$. Observe:
\begin{Lemma} \label{lem:perfbd}
Under Assumption \ref{assump:lowpass}. Suppose that {\sf (i)} ${\bm c}_{\sf eig}^\top {\bm B} {\bm q}_1 \neq 0$, where ${\bm q}_1$ is the top right singular vector of ${\cal H}( {\bm A} ) {\bm B}$, and {\sf (ii)} there exists $\delta >0$ such that 
\beq \label{eq:gap_cond}
\delta := \lambda_1(\overline{\bm C}_y ) - \lambda_2(\overline{\bm C}_y) - || \bm{\Delta} || >0,
\eeq
where $\lambda_i(\overline{\bm C}_y)$ is the $i$th largest eigenvalue of $\overline{\bm C}_y$.
It holds that
\beq \label{eq:perf_bd}
\begin{split}
& \| {\bm c}_{\sf eig} -  \widehat{\bm v}_1 \| \leq \sqrt{2} \eta \cdot \frac{ \| {\bm V}_{n-1}^\top {\bm B} {\bm q}_1 \| }{ |{\bm v}_1^\top {\bm B} {\bm q}_1| } + \frac{|| \bm{\Delta} ||}{\delta},
\end{split}
\eeq
where ${\bm V}_{n-1}$ collects the eigenvectors of ${\bm A}$ except for ${\bm v}_1$.
\end{Lemma}
The proof, which can be found in Appendix~\ref{sec:lemma1}, is based on the Davis-Kahan theorem. Note that
condition {\sf (ii)} requires the finite sample error to be bounded by the spectral gap $\overline{\bm C}_y$. 
It is well known that $\| \bm{\Delta} \| = {\cal O}( \sigma^2 + 1/\sqrt{m} )$ with high probability \cite{vershynin2018high}. Under \Cref{assump:lowpass}, it can be shown that $\lambda_1( \overline{\bm C}_y ) - \lambda_2 ( \overline{\bm C}_y ) = \Omega( |h(\lambda_1)|^2 ( 1 - c \eta ) )$ for some $c>0$. As such, \eqref{eq:gap_cond} holds when $m \gg 1$, $\sigma^2 \ll 1$, $\eta \approx 0$.

Eq.~\eqref{eq:perf_bd} gives an upper bound on the error in estimating the eigen-centrality via \eqref{eq:pca}. Notice that when $\widehat{\bm v}_1 \approx {\bm c}_{\sf eig}$, we can detect the top-$C$ central nodes accurately from $\widehat{\bm v}_1$. Now, the first term on the right hand side (r.h.s.) of \eqref{eq:perf_bd} depends on the low pass ratio and the correlation between ${\bm v}_1$, ${\bm B}$. Particularly, this term is small for the \emph{strong} low pass graph filter with $\eta \approx 0$, corroborating with the insight from \eqref{eq:lowpass_approx}. The second term in the r.h.s.~depends on $\| \bm{\Delta} \|$ which is controlled by the number of samples and the measurement noise. In the special case of ${\bm B} = {\bm I}$, we notice ${\bm q}_1 = {\bm v}_1 = {\bm c}_{\sf eig}$ and the first term in the r.h.s.~vanishes. The r.h.s.~of \eqref{eq:perf_bd} is simplified to $\| \bm{\Delta} \| / \delta$ where it only depends on the finite sample error and observation noise variance. Note that as ${\bm B} = {\bm I}$ implies $\EE[ {\bm x}_\ell {\bm x}_\ell^\top ] = {\bm I}$, i.e., white noise excitation, we recover the results in \cite{roddenberry2020blind,roddenberry2020blinda}.

\begin{figure}[t]
\begin{center}
    \includegraphics[width=.492\linewidth]{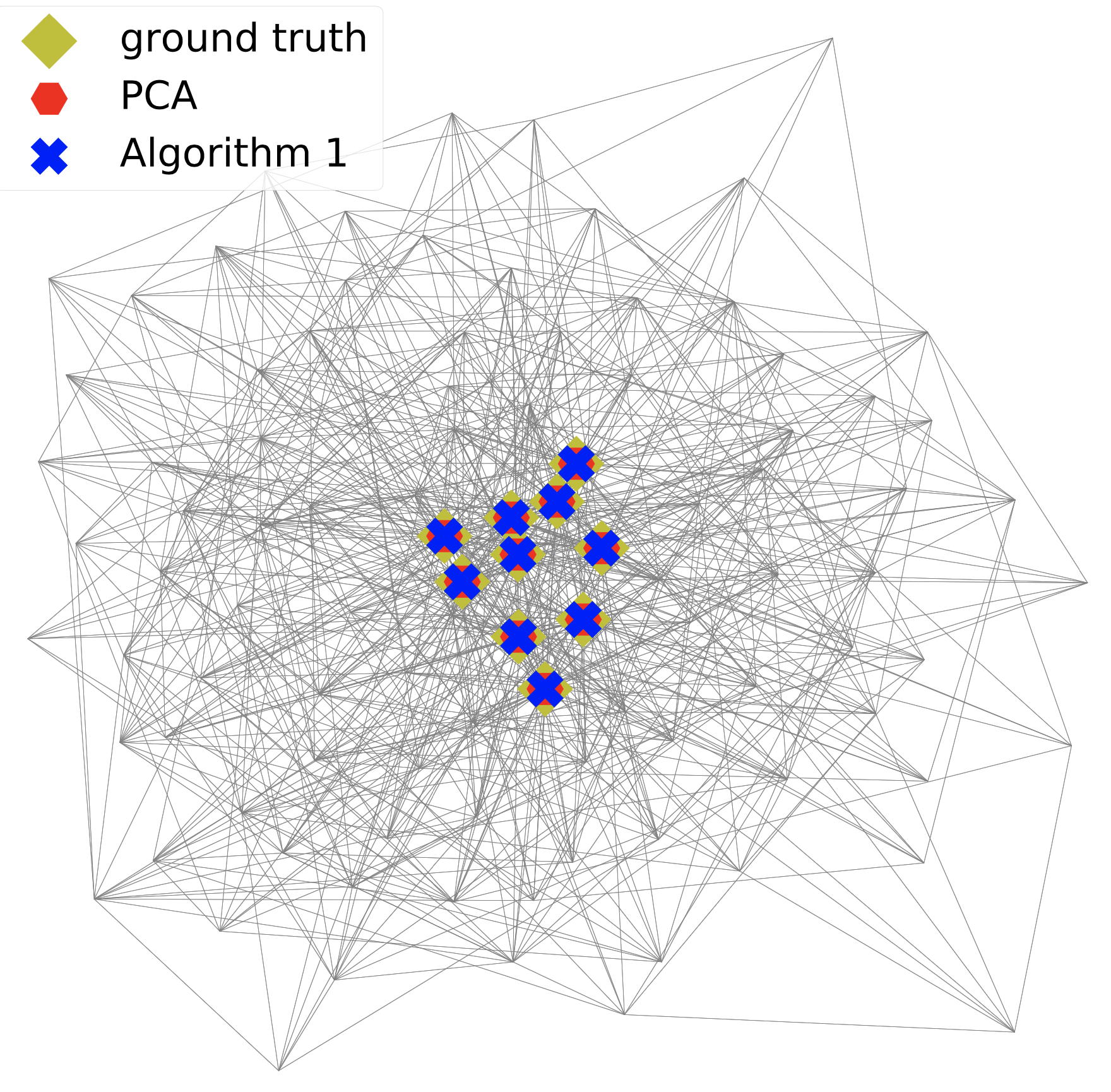}~
    \includegraphics[width=.492\linewidth]{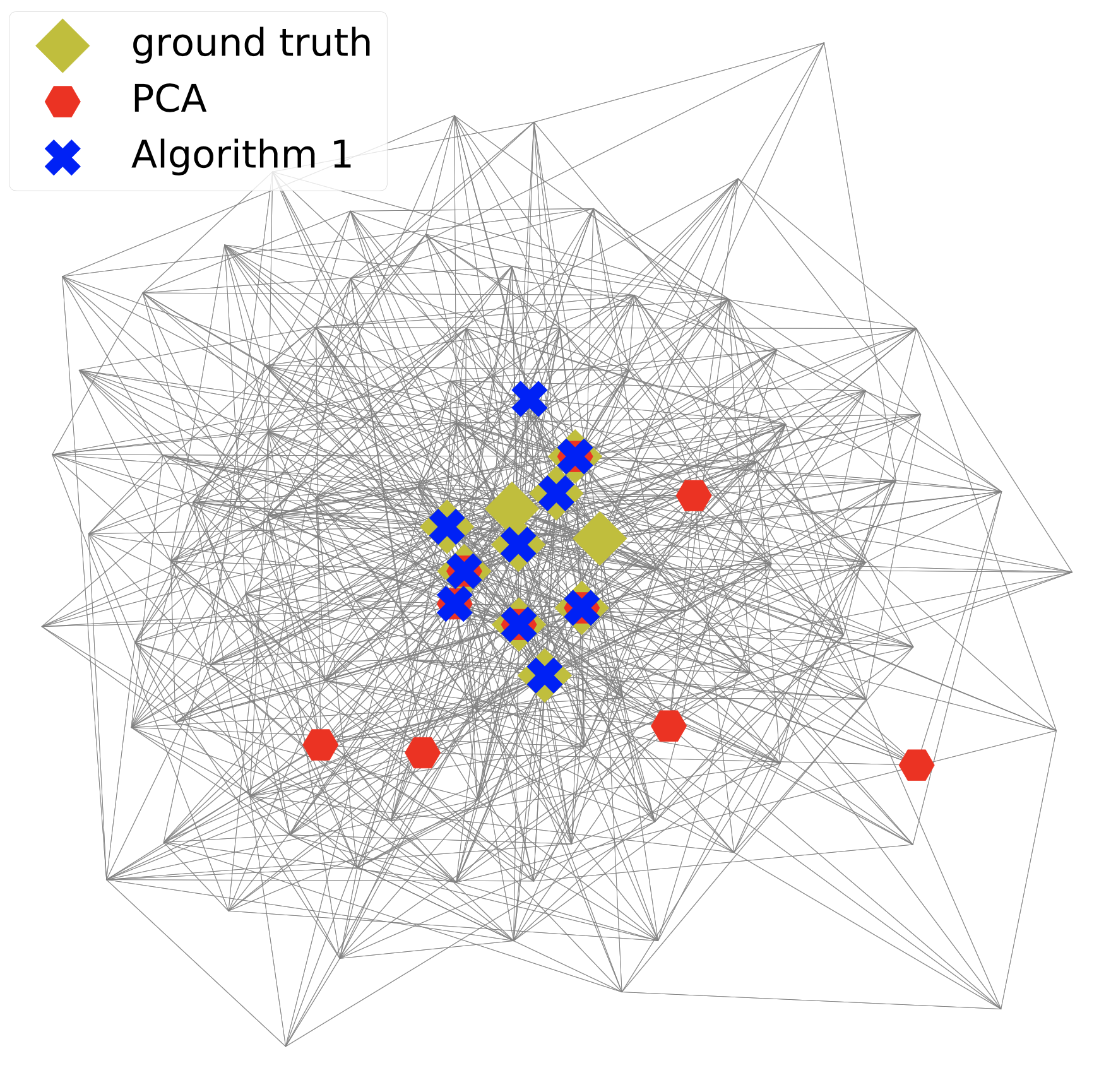}\vspace{-.2cm}
\end{center}
\caption{Toy Example illustrating the detection results of the proposed methods on a core periphery graph: (Left) with strong low pass filter ${\cal H}_{\sf strong}({\bm A}) = e^{0.1 {\bm A}}$; (Right) with weak low pass filter ${\cal H}_{\sf weak}({\bm A}) = ( {\bm I} - \frac{1}{50} {\bm A})^{-1}$.}\label{fig:reliaz}\vspace{-.4cm}
\end{figure}

To summarize, Lemma~\ref{lem:perfbd} indicates that for \emph{strong} low pass graph filter, i.e., $\eta \approx 0$, the PCA method \eqref{eq:pca} attains high accuracy for central nodes detection with $m \gg 1, \sigma^2 \ll 1$. However, when dealing with \emph{weak} low pass graph filter, i.e., $\eta \approx 1$, the method may become unreliable as shown below. 

\noindent \textbf{A Toy Example.}~~We consider detecting central nodes using the PCA method for a simple model. In Fig.~\ref{fig:reliaz}, we show the detection results for an instance of applying \eqref{eq:pca} when the observed graph signals are filtered by different low pass filters. The underlying graph topology is a $100$-nodes core-periphery graph generated from a stochastic block model with $C=10$ fully connected central nodes and the excitation is low rank with $k=40$; see \Cref{sec:synexp} for details of the generation model. We observe that the top-10 central nodes are correctly detected by PCA when the underlying graph filter is strong low pass (left panel), yet PCA misidentified a number of nodes when underlying graph filter is weak low pass (right panel).

{Before concluding this section, we remark that \Cref{lem:perfbd} can be specialized to analyze the top left singular vector of ${\cal H}({\bm A}){\bm B}$ for any low pass graph filter ${\cal H}({\bm A})$ and basis matrix ${\bm B}$. The result is given by the following corollary: 
\begin{Corollary} \label{cor:perf}
Let $\widehat{\bm v}_1, {\bm q}_1$ be the top left, right singular vector of ${\cal H}({\bm A}) {\bm B}$, respectively, where ${\cal H}({\bm A})$ is a graph filter satisfying \Cref{assump:lowpass}. Further, assume that $\widehat{\bm v}_1^\top {\bm c}_{\sf eig} \geq 0$ and ${\bm c}_{\sf eig}^\top {\bm B} {\bm q}_1 \neq 0$. It holds \vspace{-.1cm}
\beq
\| \widehat{\bm v}_1 - {\bm c}_{\sf eig} \| \leq \sqrt{2} \eta \cdot \| {\bm V}_{n-1}^\top {\bm B} {\bm q}_1 \| / | {\bm v}_1^\top {\bm B} {\bm q}_1 | = {\cal O}(\eta).\vspace{-.1cm}
\eeq
\end{Corollary}
The corollary directly relates the estimation error for the eigen-centrality to the low pass ratio. As $\eta \approx 0$, the singular value decomposition of ${\cal H}({\bm A}) {\bm B}$ generates an accurate estimate of eigen-centrality. We shall exploit this observation next. 
}\vspace{-.3cm}

\section{Detection with General Low Pass Filter}\vspace{-.1cm} \label{sec:weak}
{Consider the case of a general low pass graph filter where only $\eta < 1$ is needed in \Cref{assump:lowpass}.
This case includes weak $1$-low pass graph filters with $\eta \approx 1$ that are found in some practical models, e.g., the examples in \Cref{sec:case} have ${\cal H}({\bm A}) = ( {\bm I} - c {\bm A})^{-1}$ with $\eta \approx 1$. 
The relaxed condition has made the blind central nodes detection problem significantly more challenging. 
For instance, simple methods such as PCA may no longer work as observed from \eqref{eq:y_cov}, where the top eigenvector of ${\rm Cov}( {\bm y}_\ell )$ can be different from ${\bm c}_{\sf eig}$.

To handle this challenging case, a blind central nodes detection method is developed using two novel observations: {\sf (A)} We derive an intrinsic decomposition for the graph signal model featuring a \emph{boosted} graph filter with improved low pass ratio \cite{wai2019blind}. We then treat {\blue $\{ {\bm y}_i \}_{i=1}^m$} using a factor analysis model and demonstrate that the desired eigen-centrality vector is embedded with additive and multiplicative perturbations. {\sf (B)} By concentrating on a special case of the graph signal model with sparse and non-negative basis for the excitation, we treat a sparse non-negative matrix factorization (NMF) problem which can uniquely identify the latent parameter matrix if the unknown matrices satisfy a sufficiently scattered condition, i.e., removing the multiplicative perturbation. Then, a RPCA method is applied to remove the additive perturbation.
These observations will be discussed in order.

\noindent \textbf{Structured Factor Analysis Model.}~~
Let $\rho \geq 0$ be a free parameter\footnote{The parameter $\rho$ is introduced only for the modeling purpose and is not required by our algorithm.} and assume that ${\bm w}_\ell = {\bm 0}$ for ease of exposition. Using the matrix notation ${\bm Y} = [ {\bm y}_1 ~\cdots~{\bm y}_m ]$, ${\bm Z} = [ {\bm z}_1~\cdots~{\bm z}_m ]$, we first observe the following intrinsic decomposition from \eqref{eq:y=Hx}, \eqref{eq:x=Bz} for the graph signal matrix ${\bm Y}$:
\beq \label{eq:Y_dec}
{\bm Y} = \big\{ \tilde{\cal H}_\rho ({\bm A}) + \rho {\bm I} \big\} {\bm B} {\bm Z} = \big\{ \tilde{\cal H}_\rho ({\bm A}) {\bm B} + \rho {\bm B} \big\} {\bm Z} ,
\eeq
where $\widetilde{\cal H}_\rho ( {\bm A} )$ is a \emph{boosted graph filter} defined as \cite{wai2019blind}:
\beq \label{eq:boost}
\widetilde{\cal H}_\rho ( {\bm A} ) \eqdef {\cal H}({\bm A}) - \rho {\bm I}.
\eeq
The filter has a frequency response of $\tilde{h}_\rho( \lambda ) \eqdef h(\lambda) - \rho$. 
A useful result from \cite[Observation 1]{wai2019blind} is that under mild conditions, there exists $\rho > 0$ such that $\widetilde{\cal H}_\rho ( {\bm A} )$ is a \emph{stronger low pass graph filter} than ${\cal H}({\bm A})$ with a strictly smaller low pass ratio\footnote{Notice that \cite{wai2019blind} focused on low pass graph filters that are defined using the Laplacian matrix, yet the observation can be easily extended to the case of adjacency matrix considered in this paper.}. For example, consider the IIR graph filter ${\cal H}({\bm A}) = ( {\bm I} - c {\bm A})^{-1}$. It can be shown that $\tilde{\cal H}_\rho({\bm A})$ has an improved low pass ratio, $\tilde{\eta}$, that is bounded by
\beq
\begin{split}
\tilde{\eta} & := \min_{\rho > 0} \frac{ \displaystyle \max_{j=2,...,n} | \tilde{h}_\rho (\lambda_j)|}{ | \tilde{h}_\rho (\lambda_1)| } 
\leq \frac{\lambda_2}{\lambda_1}  \frac{ (1-c\lambda_2)^{-1}}{ (1-c\lambda_1)^{-1}}=  \frac{\lambda_2}{\lambda_1} \eta, 
\end{split}
\eeq
where the bound is obtained by setting $\rho = 1$. Together with the condition that $\lambda_1 \gg \lambda_2$ as the graph $G$ admits a core-periphery structure, we have $\tilde{\eta} \ll \eta < 1$.  
Denote the top left singular vector of $\widetilde{\cal H}_\rho ( {\bm A} ) {\bm B}$ as $\widehat{\bm v}_{\sf ideal}$. \Cref{cor:perf} shows that
\beq
\| \widehat{\bm v}_{\sf ideal} - {\bm c}_{\sf eig} \| = {\cal O}( \tilde{\eta} ).
\eeq
In other words, if the matrix $\widetilde{\cal H}_\rho ( {\bm A} ) {\bm B}$ is known, then the eigen-centrality vector can be estimated by computing a singular value decomposition (SVD). 

The main question is \emph{how to estimate the matrix $\widetilde{\cal H}_\rho ( {\bm A} ) {\bm B}$ from ${\bm Y}$?} Notice that as $\tilde{\eta} \ll 1$, $\widetilde{\cal H}_\rho ( {\bm A} ) {\bm B}$ is approximately rank-one. From \eqref{eq:Y_dec}, we observe that this low-rank matrix is embedded in ${\bm Y}$ with unknown additive ($\rho {\bm B}$) and multiplicative perturbations (${\bm Z}$), which makes it difficult to estimate. Therefore, retrieving $\widetilde{\cal H}_\rho ( {\bm A} ) {\bm B}$ requires one to leverage additional structures in the graph signals. We are inspired by Examples~\ref{ex:op} and \ref{ex:stock} to concentrate on a case with:\vspace{-.1cm}
\begin{assumption} \label{ass:nonneg}
The basis matrix ${\bm B}$ is {sparse} 
and {non-negative}, and the excitation parameters ${\bm Z}$ are {non-negative}.\vspace{-.1cm}\end{assumption}
Besides Examples~\ref{ex:op} and \ref{ex:stock}, we remark that the above can be satisfied in other applications such as in pricing experiments when the number of controllable agents is small \cite{wai2019blind}. 
Finally, we conclude that ${\bm Y}$ can be described by a \emph{structured factor} model consisting of a \emph{low-rank plus sparse} factor $\tilde{\cal H}_\rho ({\bm A}) {\bm B} + \rho {\bm B}$, and a non-negative factor ${\bm Z}$. 

To grasp an idea of how \Cref{ass:nonneg} and \eqref{eq:Y_dec} can contribute to tackling the blind central nodes detection problem, we may take a slight detour by considering the case when ${\bm Z}$ is known. 
Here, a standard algorithm is to apply the robust PCA (RPCA) method \cite{agarwal2012noisy,candes2011robust} which leverages that $\rho {\bm B}$ is a sparse matrix. By solving a convex optimization problem [cf.~\eqref{eq:rpca}], it is possible to extract the desired low-rank component $\tilde{\cal H}_\rho ({\bm A}) {\bm B}$ from ${\bm Y} {\bm Z}^\dagger = \tilde{\cal H}_\rho ({\bm A}) {\bm B} + \rho {\bm B}$. 
As our target is to tackle the blind central nodes detection problem, we next focus on the case when ${\bm Z}$ is unknown
where the property ${\bm B}, {\bm Z} \geq {\bm 0}$ is exploited by the NMF technique.\vspace{-.2cm}

\subsection{Identifying Factors in \eqref{eq:Y_dec} using NMF}\label{sec:nmf}\vspace{-.1cm}
To fix notations, let us define
\beq \label{eq:gnd}
{\bm H}_\star = {\cal H}({\bm A}) {\bm B} = \tilde{\cal H}_\rho ({\bm A}) {\bm B} + \rho {\bm B}, ~~ {\bm Z}_\star = {\bm Z}
\eeq
as the \emph{ground truth factors} satisfying ${\bm Y} = {\bm H}_\star {\bm Z}_\star$. We concentrate on the special case where ${\bm H}_\star \geq {\bm 0}, {\bm Z}_\star \geq {\bm 0}$.
Notice that under \Cref{ass:nonneg}, ${\bm B}, {\bm Z}$ are non-negative matrices. Further, the condition ${\bm H}_\star \geq {\bm 0}$ is satisfied for cases such as ${\cal H}({\bm A}) = ({\bm I} - c {\bm A})^{-1}$ with $c < ||{\bm A} ||^{-1}$. 
With these constraints on \eqref{eq:Y_dec}, the NMF problem finds a pair of matrices $(\widehat{\bm H}$, $\widehat{\bm Z}) \in \Re^{n \times k} \times \Re^{k \times m}$ such that
\beq \label{eq:nmf}
{\bm Y} = {\bm H}_\star {\bm Z}_\star = \widehat{\bm H} \widehat{\bm Z},~~\widehat{\bm H} \geq {\bm 0},~~\widehat{\bm Z} \geq {\bm 0}.
\eeq
We aim to study \emph{identifiability conditions} which guarantee $\widehat{\bm Z} = {\bm Z}_\star$, $\widehat{\bm H} = {\bm H}_\star$ so that the low-rank matrix $\tilde{\cal H}_\rho({\bm A}) {\bm B}$ can be derived from the latter, e.g., by applying RPCA.
We shall emphasize that the non-negativity constraints are important. Notably, in their absence, for any invertible ${\bm Q} \in \RR^{k \times k}$, the pair $({\bm H}_\star {\bm Q}, {\bm Q}^{-1} {\bm Z}_\star)$ is admissible for the model in \eqref{eq:nmf}. Such ambiguity forbids us from estimating $\tilde{\cal H}_\rho ({\bm A}) {\bm B}$ with methods like RPCA since the matrix ${\bm H}_\star {\bm Q} = \tilde{\cal H}_\rho( {\bm A}) {\bm B} {\bm Q} + \rho {\bm B} {\bm Q}$ does not admit a low-rank plus sparse decomposition. 
It is unclear if the desired matrix $\tilde{\cal H}_\rho( {\bm A}) {\bm B} {\bm Q}$ can be extracted.

A major benefit of  NMF is that under a relatively mild condition, \eqref{eq:nmf} admits an \emph{essentially unique} factorization, i.e., the recovered factors are only subjected to diagonal and permutation ambiguities. We observe:
\begin{Fact}\label{fact:unique}  \cite[Theorem 4]{huang2013non}
Suppose that the rows of ${\bm H}_\star$ and columns of ${\bm Z}_\star$ are sufficiently scattered, i.e.,
\beq \label{eq:identifiable}
\{ {\bm x} \in \Re^k : {\bm x}^\top {\bf 1} \geq \sqrt{k-1} \| {\bm x} \|_2 \} \subseteq {\rm cone} \{ {\bm H}_\star^\top \},
\eeq
${\rm cone} \{ {\bm H}_\star^\top \} \not\subset {\rm cone} \{ {\bm Q} \}$ for any orthonormal matrix ${\bm Q}$ except for the permutation matrix, and the same holds for ${\bm Z}_\star$. Then, any solution $(\widehat{\bm H}$, $\widehat{\bm Z})$ satisfying \eqref{eq:nmf} can be written as 
\beq \label{eq:unique}
(\widehat{\bm H}, \widehat{\bm Z}) = ( {\bm H}_\star {\bm D} \bm{\Pi}, \bm{\Pi}^\top {\bm D}^{-1} {\bm Z}_\star ),
\eeq
where ${\bm D}$ is a positive diagonal matrix, and $\bm{\Pi}$ is a permutation matrix.\vspace{-.2cm}
\end{Fact}
The sufficiently scattered conditions are satisfied when the ground truth factors are mildly sparse \cite{huang2013non}. Our next endeavor is to verify these conditions in terms of the graph signal properties and their impact on the performance of blind central nodes detection. 
We proceed by discussing two issues in order. 

\noindent \textbf{{\sf 1)} Identifiability.}~ The first issue is whether the identifiability conditions in \Cref{fact:unique} are satisfied for the filtered graph signals in \eqref{eq:gnd}. Unfortunately, at the first glance this appears to be impossible since a necessary condition for \Cref{fact:unique} to hold is that ${\bm H}_\star$ contains at least $k-1$ zeros at every row \cite[Corollary 2]{huang2013non}, while ${\bm H}_\star = {\cal H}({\bm A}) {\bm B}$ is a \emph{dense} matrix in general. 

As a remedy, we consider the approximate model $ {\bm Y} \approx \rho {\bm B} {\bm Z} $ and exploit that the basis matrix ${\bm B}$ is sparse. Interestingly, the next result recognizes that ${\bm H}_\star \approx \rho {\bm B}$ when the original graph filter' low pass ratio satisfies $\eta \approx 1$, i.e., the graph filter is weak low pass. To derive this result, we notice the constant $\rho$ is a parameter that can be freely adjusted as it is implicit in the signal model. Specific to the current context, we seek for $\rho > 0$ such that the ratio of $\| \widetilde{\cal H}_\rho( {\bm A}) {\bm B} \|$ to $\rho \|{\bm B} \|$ is minimized. We observe that:\vspace{-.1cm}
\begin{Lemma} \label{lem:sparse}
Under Assumption~\ref{assump:lowpass}. Suppose that the frequency response $h(\lambda)$ is a convex function and is non-negative over $[\lambda_n,\lambda_1]$. 
Define $h_{\sf min} := \min_{i=2,...,n} h(\lambda_i)$ as the smallest frequency response. It holds  
\beq \label{eq:sparse_ratio}
\min_{\rho > 0} \frac{ \| \widetilde{\cal H}_\rho( {\bm A}) {\bm B} \| }{ \rho \| {\bm B} \| } \leq \frac{ 1 - \eta + \frac{ \max\{h(\lambda_2),h(\lambda_n)\} - h_{\sf min} }{h(\lambda_1)} }{ 1 + \eta - \frac{ \max\{h(\lambda_2),h(\lambda_n)\} - h_{\sf min} }{h(\lambda_1)} }. 
\eeq
Furthermore, if $\max\{h(\lambda_2),h(\lambda_n)\} - h_{\sf min} \ll h(\lambda_1)$, then:
\beq
\min_{\rho > 0} \frac{ \| \widetilde{\cal H}_\rho( {\bm A}) {\bm B} \| }{ \rho \| {\bm B} \| } \lesssim \frac{ 1 - \eta }{ 1 + \eta }. \vspace{-.1cm}
\eeq
\end{Lemma}
The proof can be found in Appendix~\ref{app:sparse} and is based on bounding $\rho$ which minimizes the left hand side of \eqref{eq:sparse_ratio}. 
In particular, the above shows that \emph{the approximation ${\bm H}_\star \approx \rho {\bm B}$ holds if $\eta \approx 1$}.
Note that the convexity assumption on $h(\cdot)$ is satisfied by the graph filters considered in this work.

Contrary to ${\bm H}_\star$, the matrix $\rho {\bm B}$ is sparse and non-negative. Now consider \Cref{fact:unique}, while the sufficiently scattered condition \eqref{eq:identifiable} is NP-complete to check, it is noted ``\emph{if the latent factors are sparse, it is more likely that the sufficient condition given in \Cref{fact:unique} is satisfied}'' \cite{huang2013non}. For an NMF model with $\widehat{\bm Y} = \rho {\bm B} {\bm Z}$, a special case where the stated conditions hold is when ${\rm cone} \{ {\bm e}_1, ..., {\bm e}_k \} \subseteq {\rm cone} \{ {\bm B}^\top \}$, i.e., ${\bm B}$ contains $k$ pure pixel row vectors. When combined with \Cref{lem:sparse}, we observe:
\begin{Corollary} \label{cor:idd}
Assume the system parameters $({\bm B}, {\bm Z})$ satisfy \Cref{ass:nonneg} and the sufficiently scattered condition in \Cref{fact:unique}, and the assumptions in \Cref{lem:sparse} are satisfied with $\eta \approx 1$. Then solving \eqref{eq:nmf} yields 
\beq 
\widehat{\bm Z} \approx \bm{\Pi}^\top {\bm D}^{-1} {\bm Z}_\star,
\eeq
for some positive diagonal and permutation matrices ${\bm D}$, $\bm{\Pi}$.\vspace{-.1cm}
\end{Corollary}
It has been demonstrated in \cite{huang2013non, fu2019nonnegative} that the sufficiently scattered conditions hold as long as $({\bm B}, {\bm Z})$ are not fully dense. 
Concretely, when ${\bm B}$ is a random sparse matrix, the NMF model is shown numerically to be identifiable with high probability when the ratio $n/k$ is large.\vspace{.1cm}

\noindent \textbf{{\sf 2)} Ambiguities.}~ Having verified the identifiability conditions for the NMF model, we study the effects on central nodes detection performance caused by diagonal and permutation ambiguities. We find that these ambiguities are typically dismissed in existing applications of NMF such as source separation, topic modeling, and clustering, etc., as they do not directly affect the performances in these applications. 

For the blind central nodes detection problem, obtaining $( \widehat{\bm B}, \widehat{\bm Z} )$ is not the final goal. 
Instead, our objective is to estimate the eigen-centrality. As \Cref{cor:idd} guarantees that $\widehat{\bm Z} \approx \bm{\Pi}^\top {\bm D}^{-1} {\bm Z}_\star$, we apply the RPCA method to treat ${\bm Y} \widehat{\bm Z}^\dagger \approx {\bm H}_\star {\bm D} \bm{\Pi}$ and recover the low-rank matrix therein as
\beq \label{eq:rpca_approx}
\widehat{\bm L} = \tilde{\cal H}_\rho({\bm A}) {\bm B} {\bm D} \bm{\Pi}.
\eeq 
Here, ${\bm D}$ can be any positive diagonal matrix, the top left singular vector of $\widehat{\bm L}$ is a poor estimate of the eigen-centrality.  

Our remedy is to estimate the row sum ${\bm Z}_\star {\bf 1}$ a-priori in order to fix the diagonal ambiguity matrix ${\bm D}$. Consider\vspace{-.1cm}
\begin{assumption} \label{ass:const}
For any $\ell, j$, $z_{\ell,j}$ is an independent random variable with $\EE[ z_{\ell,j} ] = \alpha$ and sub-Gaussian parameter $\sigma_z$.\vspace{-.1cm}
\end{assumption}
The above condition is related to having persistent excitation for the external sources. 
Together with the approximation analyzed in \Cref{lem:sparse}, we study a modified NMF model from \eqref{eq:nmf} with additional constraint as:
\beq \label{eq:nmf_fc}
{\bm Y} = \rho {\bm B} {\bm Z} + {\bm W} \equiv \widehat{\bm B} \widehat{\bm Z} + {\bm W},~\widehat{\bm B} \geq {\bm 0},~\widehat{\bm Z} \geq {\bm 0},~\widehat{\bm Z}{\bf 1} = {\bf 1},
\eeq
where ${\bm W}$ is the approximation error from ${\bm Y} \approx \rho {\bm B} {\bm Z}$. 
We analyze the low-rank matrix recovered using the RPCA method from $\widehat{\bm Z}$ satisfying \eqref{eq:nmf_fc} and obtain
the lemma below:
\begin{Lemma} \label{cor:diagonal}
Under Assumptions~\ref{assump:lowpass} and \ref{ass:const}. Let $\widehat{\bm L}$ be defined in \eqref{eq:rpca_approx} using $\widehat{\bm Z}$ computed from \eqref{eq:nmf_fc}, and its top left (resp.~right) singular vector be $\widetilde{\bm v}_1$ (resp.~${\bm q}_1$).
For any $ \delta >0$ and sufficiently large $m$, it holds with probability at least $1- 2k \delta$,
\beq \label{eq:lem3} 
\hspace{-.2cm} \| \widetilde{\bm v}_1 - {\bm c}_{\sf eig} \|_2 \leq \sqrt{2} \, \tilde{\eta} \, \frac{  \alpha + \frac{2 \sigma_z^2}{m} \log(\delta^{-1}) }{\alpha - \frac{2 \sigma_z^2}{m} \log(\delta^{-1}) } \frac{ \| {\bm V}_{n-1}^\top {\bm B} \bm{\Pi} {\bm q}_1 \|_2 }{ |{\bm v}_1^\top {\bm B} \bm{\Pi} {\bm q}_1| }.\vspace{-.2cm}
\eeq
\end{Lemma} 
The proof, which is based on the Hoeffding's inequality and \Cref{cor:perf}, can be found in Appendix~\ref{app:diagonal}. As the number of samples grows $m \rightarrow \infty$, the estimate of the eigen-centrality vector becomes more accurate.
We remark that \Cref{ass:const} can be relaxed to allow for heterogeneous mean in ${\bm z}_\ell$, i.e., $\EE[ z_{\ell,j} ] = \alpha_j$. In the latter case, the bound \eqref{eq:lem3} will be relaxed to ${\cal O}( \tilde{\eta} \, \max_{i,j} \frac{\alpha_i}{\alpha_j}) $ when $m \gg 1$. \vspace{-.2cm}
}

\algsetup{indent=1em}
\begin{algorithm}[tb]
	\caption{Two-Stage Decomposition Algorithm}\label{alg:mega}
	\begin{algorithmic}[1]
		\STATE {\textbf{INPUT}}: graph signal matrix ${\bm Y}$, regularization parameters $\lambda_B, \lambda_S, \lambda_L > 0$, and the dimension of latent excitation parameters $k$.
		\STATE \label{line:nmf} Tackle the NMF problem \eqref{eq:sparsedl} to estimate the latent excitation parameters $\widehat{\bm Z}$; see \Cref{alg:nmf}.
		\STATE Solve the RPCA problem \eqref{eq:rpca} using $\widehat{\bm H}$ from \eqref{eq:hhat_main}.
		\STATE \label{step:svd} Compute $\widetilde{\bm v}_1$ as the top left singular vector of $\widehat{\bm L}$. 
		\STATE {\textbf{OUTPUT}}: select $C$ elements with the highest magnitude in  $\widetilde{\bm v}_1$ as the detected central nodes.
	\end{algorithmic} 
\end{algorithm}

\vspace{-.2cm}
\subsection{Two-stage Decomposition Algorithm}\vspace{-.1cm}
{The previous discussions established the theoretical foundation of detecting central nodes from graph signals filtered by a general low pass graph filter. Hereafter, we describe how to utilize these observations to develop a practical algorithm.

In light of the structured factor model \eqref{eq:Y_dec} and our theoretical findings, we split the detection problem into \emph{two stages} --- the first stage estimates the latent excitation parameters $\widehat{\bm Z} \approx {\bm Z}_\star$; the second stage estimates the low-rank matrix $\tilde{\cal H}_\rho({\bm A}) {\bm B} {\bm D} \bm{\Pi}$ from ${\bm Y}$ based on $\widehat{\bm Z}$. 
Finally, the central nodes are detected through computing and ranking the magnitude of top left singular vector of the estimated $\tilde{\cal H}_\rho({\bm A}) {\bm B} {\bm D} \bm{\Pi}$. 

The first stage is the most challenging as it relies on identifying the factors from the NMF model \eqref{eq:nmf_fc}. In light of Assumptions \ref{ass:nonneg} and \ref{ass:const}, we seek for a solution where the `left' factor is a sparse matrix and the row sums of the `right' factor equal to one. Taking into account the noisy observations and the approximation errors, we consider a sparse NMF criterion:
\beq \label{eq:sparsedl}
\begin{array}{rl}
\ds \min_{ \widehat{\bm B} \in \Re^{n \times k}, \widehat{\bm Z} \in \Re^{k \times m} } & {\textstyle \frac{1}{2}} \| {\bm Y} - \widehat{\bm B} \widehat{\bm Z} \|_F^2 + \lambda_B \| {\rm vec}( \widehat{\bm B} ) \|_1 \\
\text{s.t.} & \widehat{\bm Z} {\bf 1} = {\bf 1},~\widehat{\bm B} \geq {\bm 0},~ \widehat{\bm Z} \geq {\bm 0},
\end{array}
\eeq
such that $\lambda_B > 0$ is a regularization parameter promoting sparsity for the solution $\widehat{\bm B}$.
Notice that while sparse NMF has been applied in empirical studies \cite{hoyer2002non}, our criterion is directly motivated by the identifiability conditions in \Cref{cor:idd} for the graph signals as we aim to recover $\widehat{\bm B}, \widehat{\bm Z}$. The above problem can be efficiently tackled as a customized solver for \eqref{eq:sparsedl} will be discussed in \Cref{sec:practical}.

An important observation is when the sufficiently scattered conditions hold for the matrices $({\bm B}, {\bm Z})$, \Cref{cor:idd} only shows that the solution of \eqref{eq:sparsedl} satisfies $\widehat{\bm Z} \approx \bm{\Pi}^\top {\bm D}^{-1} {\bm Z}_\star$. However, our target is to estimate ${\bm H}_\star$ instead. The latter is achieved by solving the least square problem, e.g., 
\beq \label{eq:hhat_main} \textstyle
\min_{ \widehat{\bm H} \in \RR^{n \times k} }~\| {\bm Y} - \widehat{\bm H} \widehat{\bm Z} \|_F^2.
\eeq

Notice that the solution to \eqref{eq:hhat_main}, i.e., $\widehat{\bm H} \approx \tilde{\cal H}_\rho({\bm A}) {\bm B} {\bm D} \bm{\Pi} + \rho {\bm B} {\bm D} \bm{\Pi}$, is composed of a low-rank matrix and a sparse matrix. As such, in the second stage of our algorithm, we solve the RPCA problem:
\begin{equation}\label{eq:rpca}
\min_{ \widehat{\bm L}, \widehat{\bm S} \in \RR^{n \times k} }~\| \widehat{{\bm H}} - \widehat{\bm L} - \widehat{\bm S} \|_F^2 + \lambda_L \| \widehat{\bm L} \|_\star + \lambda_S \| {\rm vec}(\widehat{\bm S}) \|_1,
\end{equation}
where $\lambda_S, \lambda_L > 0$ are regularization parameters, to obtain an estimate of $\tilde{\cal H}_\rho({\bm A}) {\bm B} {\bm D} \bm{\Pi}$.
Similar to \cite{wai2019blind}, this step may also be directly applied when ${\bm Z}$ is known [cf.~Fig.~\ref{fig:overview}]. 
With appropriate parameters $\lambda_S, \lambda_L$, the solution to \eqref{eq:rpca} satisfies $\widehat{\bm L} \approx \tilde{\cal H}_\rho({\bm A}) {\bm B} {\bm D} \bm{\Pi}$. The central nodes can be detected subsequently by computing the top singular vector of the former matrix. We remark that \eqref{eq:rpca} is a convex problem that can be solved efficiently using existing algorithms, e.g., \cite{aravkin2014variational}. 

Finally, our method for blind central nodes detection is summarized in \Cref{alg:mega}. 
Notice that \Cref{alg:mega} requires knowing the excitation's rank $k$ which can be estimated heuristically from observing the rank of ${\rm Cov}( {\bm y}_\ell )$.

\noindent \textbf{Discussion.}~~We briefly discuss the performance of \Cref{alg:mega}. 
For the first stage, observe that \eqref{eq:sparsedl} aims at estimating ${\bm Z}_\star$ by fitting a pair $(\widehat{\bm B}, \widehat{\bm Z})$ to ${\bm Y} \approx \rho {\bm B} {\bm Z}$. 
To this end, \Cref{cor:idd} shows that the estimation quality  depends on two factors: {\sf (i)} whether the matrices ${\bm B}, {\bm Z}$ satisfy the sufficiently scattered conditions, and {\sf (ii)} whether the graph filter is weak low pass [cf.~\Cref{lem:sparse}]. 

For the second stage, it is known that if the matrix $\widehat{\bm H}$ admits a low-rank plus sparse structure, then solving the RPCA problem \eqref{eq:rpca} decomposes $\widehat{\bm H}$ accordingly to the desired low-rank and sparse components. Under these premises, using \cite[Corollary 1]{agarwal2012noisy}, it can be shown that the vector $\widetilde{\bm v}_1$ found in Step~\ref{step:svd} of \Cref{alg:mega} is ${\cal O}(\tilde{\eta})$-close to ${\bm c}_{\sf eig}$. With $\tilde{\eta} \ll 1$ due to boosted graph filter, the top-$C$ central nodes can be detected.

{\blue The careful readers may notice that when the original graph filter is strong low pass, i.e., $\eta \approx 0$, the approximation ${\cal H}({\bm A}) {\bm B} \approx \rho {\bm B}$ induced by \Cref{lem:sparse} is not guaranteed and the NMF model may become unidentifiable. However, it does not prevent \Cref{alg:mega} from correctly detecting the central nodes.  This is because the data matrix ${\bm Y}$ as well as the matrix $\widehat{\bm H}$ in step 3 of \Cref{alg:mega} are already close to rank-one since $\eta \approx 0$. 
As such, the obtained estimate $\widetilde{\bm v}_1$ remains ${\cal O}(\eta)$-close to ${\bm c}_{\sf eig}$ despite erroneous estimation of $\widehat{\bm Z}$. 
We observe from Fig.~\ref{fig:reliaz} that \Cref{alg:mega} can successfully detect the central nodes regardless of the low pass filters' strengths.
Therefore, we conclude that \Cref{alg:mega} is suitable for blind central nodes detection with \emph{general} low pass graph filter.
}}\vspace{-.2cm}

{\blue 
\begin{Remark}
Notice that Step 3 of \Cref{alg:mega} recovers the low rank matrix as $\widehat{\bm L} \approx \tilde{\cal H}_\rho({\bm A}) {\bm B} {\bm D} \bm{\Pi}$. The latter can be treated as a set of graph signals filtered by a strong low pass graph filter $\tilde{\cal H}_\rho({\bm A})$ with the excitation $[\tilde{\bm b}_1,\ldots, \tilde{\bm b}_K] \equiv {\bm B} {\bm D} \bm{\Pi}$. It is an interesting future direction to develop graph learning algorithms for recovering the adjacency matrix ${\bm A}$ from $\widehat{\bm L}$. 
\end{Remark}
}

{ 
\section{Practical Issues}\vspace{-.2cm} \label{sec:practical}
We dedicate this section to the practical issues in applying the proposed blind central nodes detection method. \vspace{-.2cm} 
}

\subsection{Efficient Implementation of \eqref{eq:sparsedl}}\vspace{-.2cm} \label{sec:noise}
We provide details of an efficient algorithm for tackling problem \eqref{eq:sparsedl}. Observe that the latter problem is non-convex, we apply an alternating minimization strategy based on projected gradient descent (PGD) similar to \cite{wu2020hybrid}. Let us define the following objective function:
\beq \label{eq:obj_nmf}
f( \widehat{\bm B}, \widehat{\bm Z} ) \eqdef {\textstyle \frac{1}{2}} \| {\bm Y} - \widehat{\bm B} \widehat{\bm Z} \|_F^2 + \lambda_B {\textstyle \sum_{i,j}} \widehat{B}_{ij} 
\eeq
and the simplex constraint for $\hat{\bm Z}$:
\[
\Delta^K = \{ {\bm Z} \in \Re^{k \times m} : {\bm Z} \geq {\bm 0}, {\bm Z} {\bf 1} = {\bf 1} \}.
\]
We notice that under the constraint $\widehat{\bm B} \geq {\bm 0}$, minimizing the objective function of \eqref{eq:sparsedl} is equivalent to minimizing $f( \widehat{\bm B}, \widehat{\bm Z} )$ as $\| {\rm vec}( \widehat{\bm B} ) \|_1 = \sum_{i,j} \widehat{B}_{i,j}$. The benefit of solving \eqref{eq:obj_nmf} is that the latter function is differentiable.

\algsetup{indent=1em}
\begin{algorithm}[tb]
	\caption{Alternating Projected Gradient Descent for \eqref{eq:sparsedl}}\label{alg:nmf}
	\begin{algorithmic}[1]
		\STATE {\textbf{INPUT}}: graph signals matrix ${\bm Y}$, the dimension of latent parameters $k$.
		\STATE Initialize $\widehat{\bm B}^{(0)} \geq {\bm 0}$ and $\widehat{\bm Z}^{(0)} \geq {\bm 0}$ satisfying $\widehat{\bm Z}^{(0)} {\bf 1} = {\bf 1}$.
		\FOR{$t=0,...,T-1$}
		\STATE Select step sizes $\alpha_t, \beta_t$ according to \eqref{eq:stepsize}. 
		\STATE $\widehat{\bm B}^{(t+1)} = \big( \widehat{\bm B}^{(t)} - \alpha_t \grd_{\bm B} f( \widehat{\bm B}^{(t)}, \widehat{\bm Z}^{(t)} ) \big)_+$ \label{line:b}
		\STATE $\widehat{\bm Z}^{(t+1)} = {\cal P}_{\Delta^k} \big( \widehat{\bm Z}^{(t)} - \beta_t \grd_{\bm Z} f( \widehat{\bm B}^{(t+1)}, \widehat{\bm Z}^{(t)} ) \big)$ \label{line:z}
		\ENDFOR
		\STATE {\textbf{OUTPUT}}: estimate of latent parameter matrix $\widehat{\bm Z}^{(T)}$.
	\end{algorithmic} 
\end{algorithm}

Based on this modification, the alternating PGD algorithm is described in Algorithm~\ref{alg:nmf}. The algorithm is initialized by picking random matrices $\widehat{\bm B}^{(0)}, \widehat{\bm Z}^{(0)}$, or using heuristics such as ALS \cite{cichocki2009fast}. Notice that in line~\ref{line:b}, the operator $(\cdot)_+$ denotes the element-wise maximum operator $\max\{\cdot,0\}$; in line~\ref{line:z}, the operator ${\cal P}_{\Delta^k} (\cdot)$ denotes the Euclidean projection onto the simplex constraint set. By observing the decomposition $\Delta^k = \Delta_1 \times \cdots \times \Delta_k$ where $\Delta_i$ is the $m$-dimensional probability simplex constraining the $i$th row vector, the projection can be performed efficiently by running $k$ instances of \cite[Algorithm 1]{duchi2008efficient} in parallel, {\blue each involving a complexity of ${\cal O}(m \log m)$.} 
Finally, at iteration $t$, we select the step sizes by calculating
\beq \label{eq:stepsize}
\begin{split}
\alpha_t & = \max\Big\{ \delta_B, \frac{a}{ \| (\widehat{\bm Z}^{(t)})^\top \widehat{\bm Z}^{(t)} \| } \Big\}, \\
\beta_t & = \max\Big\{ \delta_Z, \frac{b}{ \| (\widehat{\bm B}^{(t+1)})^\top \widehat{\bm B}^{(t+1)} \| } \Big\}, 
\end{split}
\eeq
where $\delta_B, \delta_Z > 0$ are small pre-fixed constants to prevent numerical instability, and $a, b \in (0,1]$ are pre-fixed step size parameters. 
The update steps in line~\ref{line:b}, \ref{line:z} may be repeated for several times to improve convergence.

It is known that Algorithm~\ref{alg:nmf} converges to a stationary solution of \eqref{eq:sparsedl} {\blue and the overall per-iteration complexity is ${\cal O}( nkm + km \log m )$}. To this end, we modify the result proven in \cite{wu2020hybrid} for general hybrid BCD algorithm, specialized to our algorithm when both variables are updated via the PGD:
\begin{Fact} \label{fact:conv}
Denote the iterates of Algorithm~\ref{alg:nmf} as ${\bm x}^{(t)} = ( \widehat{\bm B}^{(t)}, \widehat{\bm Z}^{(t)} )$. Then, for any $T \geq 1$, it holds
\beq \label{eq:convergence}
\min_{t = 0,...,T-1} \max_{ {\bm y} \in {\cal B}^0 \times \Delta^k } \langle \grd f( {\bm x}^{(t)} ) , {\bm x}^{(t)} - {\bm y} \rangle = {\cal O}(1/T),
\eeq
where ${\cal B}^0 = \{ {\bm B} : {\bm B} \geq {\bm 0}, \sum_{ij} B_{ij} \leq \lambda_S^{-1} f( \widehat{\bm B}^{(0)}, \widehat{\bm Z}^{(0)} ) \}$ is a compact subset of $\Re^{n \times k}$.
\end{Fact}
The above is proven in Appendix~\ref{app:factconv} through establishing that the objective value is monotonically decreasing, as such ${\cal B}^0 \times \Delta^k$ must contain a stationary solution to \eqref{eq:nmf_fc}.
It follows that \eqref{eq:convergence} implies Algorithm~\ref{alg:nmf} finds an $\epsilon$-stationary solution to \eqref{eq:sparsedl} in ${\cal O}(1/\epsilon)$ iterations, which is a standard convergence rate for constrained non-convex optimization.\vspace{-.2cm}

\subsection{Distinguishing Strong and General Low Pass Filter}\vspace{-.1cm}
{As overviewed in Fig.~\ref{fig:overview}, the PCA method \eqref{eq:pca} is suitable for detecting central nodes when the graph filter is strong low pass with $\eta \approx 0$; while the NMF-based \Cref{alg:mega} is suitable for general low pass graph filter with $\eta < 1$. 
As illustrated in Section~\ref{sec:exp}, \Cref{alg:mega} demonstrates good performances in all cases.
However, to balance between computation complexity and performance, it is desirable to select a suitable algorithm for the given dataset. 
As it is unknown in general if the underlying graph filter is strong low pass or not, a heuristic is to determine that the graph filter is strong low pass if the the covariance matrix ${\rm Cov}( {\bm y}_\ell )$ is approximately rank-one. We leave it as a future work to develop a data-driven algorithm to detect the type of graph filter; see \cite{zhu2020estimating, he2021identifying}.\vspace{-.2cm}}

\section{Numerical Experiments}\vspace{-.1cm}
\label{sec:exp}
In this section, we present numerical experiments on the proposed centrality nodes detection methods. We validate their efficacy compared to the state-of-the-art algorithms.\vspace{-.2cm}

\subsection{Experiments on Synthetic Data}\vspace{-.1cm} \label{sec:synexp}
We evaluate the performances of our algorithms using synthetic graph signals. 
{Two types of random graph models are considered. The first random graph is the \emph{core-periphery ({\sf CP}) graph} \cite{cucuringu2016detection} described by a stochastic block model with $2$ blocks and $n$ nodes. The node set ${\cal V} = \{1,...,n\}$ is partitioned into ${\cal V}_{\sf c} = \{1,...,10\}$ and ${\cal V}_{\sf p} = {\cal V} \setminus {\cal V}_{\sf c}$. For any $i,j \in {\cal V}$, an edge is assigned independently with probability $p_1$ if $i,j \in {\cal V}_{\sf c}$; with probability $\min\{ p_1, 4p_2 \}$ if $i \in {\cal V}_{\sf c}, j \in {\cal V}_{\sf p}$; and with probability $p_2$ if $i,j \in {\cal V}_{\sf p}$. Assume that $p_1 \gg p_2$, the nodes in ${\cal V}_{\sf c}$ are connected with a higher density of edges than the other nodes.
The second random graph is the \emph{Barabasi-Albert ({\sf BA}) model} \cite{newman2018networks}. The model is constructed according to a preferential attachment mechanism: during the graph generation, every new node is connected to $m_{\sf BA} = 10$ existing nodes selected randomly proportional to their degrees. 
Again, the constructed graph exhibits a core-periphery structure as the oldest nodes have high degrees. For the experiments with {\sf BA} graph, the ground truth top-$C$ central nodes, ${\cal V}_{\sf c}$, are computed from evaluating the eigen-centrality vector ${\bm c}_{\sf eig}$. 
}

We generate the basis matrix ${\bm B}\in\Re^{n\times k}$ as a sparse matrix such that $B_{ij} = M_{ij} \widetilde{B}_{ij}$, where $M_{ij}, \widetilde{B}_{ij}$ are independent r.v.s, $M_{ij} \in \{0,1\}$ is Bernouli  with $\EE[ M_{ij} ] = 0.1$, and $\widetilde{B}_{ij} \sim {\cal U}([0.1,1])$, i.e., $10\%$ of the entries in ${\bm B}$ are non-zero. For the latent parameter matrix ${\bm Z}\in\Re^{k\times m}$, we have $Z_{ij} = N_{ij} \widetilde{Z}_{ij}$, where $N_{ij}, \widetilde{Z}_{ij}$ are independent r.v.s, $N_{ij} \in \{0,1\}$ is Bernouli with $\EE[ N_{ij} ] = 0.6$, and $\widetilde{Z}_{ij} \sim {\cal U}([0.1,1])$, i.e., $60\%$ of the entries in ${\bm Z}$ are non-zero. 

The observed graph signal is generated as \eqref{eq:y=Hx}, \eqref{eq:x=Bz} with noise variance $\sigma^2 = 0.01$. 
We consider two low pass graph filters: {\sf (a)} ${\cal H}_{\sf weak}( {\bm A}) = ( {\bm I} - \frac{1}{50} {\bm A} )^{-1}$, {\sf (b)} ${\cal H}_{\sf strong}({\bm A}) = e^{0.1 {\bm A}}$. Notice that ${\cal H}_{\sf weak}( {\bm A})$ is \emph{weak} low pass with $\eta \approx 1$, meanwhile ${\cal H}_{\sf weak}({\bm A})$ is \emph{strong} low pass with $\eta \ll 1$. 
Fix any $C \in \NN$, we focus on the performance of central nodes detection by computing the \emph{error rate} of detecting the nodes in ${\cal V}_{\sf c}$ as the top-$C$ central nodes. 
Specifically, let $\widehat{\cal V}_{\sf c}$ be the set of central nodes detected by the algorithms. We define the error rate as
\beq \textstyle
{\sf Error~rate} = \EE \big[ \frac{1}{C} | {\cal V}_{\sf c} \cap \widehat{\cal V}_{\sf c} | \big],
\eeq 
and perform 100 Monte-carlo trials to approximate the above.

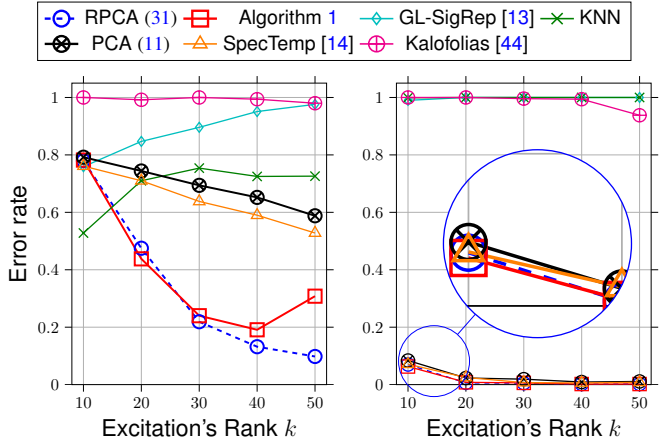
\begin{figure}[t]
    \centering
    {\sf 
    \resizebox{.9\linewidth}{!}{\input{legend_tikz.tex}}\\[.2cm]
    \resizebox{!}{.55\linewidth}{\input{Evsk_wlp_cp}}
    \resizebox{!}{.55\linewidth}{\input{Evsk_slp_cp}}
    }\vspace{-0.0cm}
    \caption{Error rate of the central nodes detection methods against the excitation's rank $k$ on (Left) ${\cal H}_{\sf weak}({\bm A})$ and (Right) ${\cal H}_{\sf strong}({\bm A})$ for {\sf CP} graph with $p_1 = 0.4, p_2 = 0.05$ and $n=100$ nodes.  The RPCA method is plotted with a dashed line to indicate that it requires the extra input ${\bm Z}$. }\label{fig:Evsk}\vspace{-.2cm}
    \end{figure}

We benchmark the two proposed methods -- {\sf (i)} {PCA} method \eqref{eq:pca} and {\sf (ii)} {\Cref{alg:mega}}. Moreover, we test the {RPCA} method \eqref{eq:rpca} which assumes ${\bm Z}$ is known for benchmarking purpose. For \eqref{eq:rpca}, we set the regularization parameters as $\lambda_L = 0.2$, $\lambda_S = 0.2 + \frac{2}{\sqrt{k}}$ and the convex optimization problem is solved using \texttt{cvxpy} with the built-in solver \texttt{SCS}; while we set $\lambda_B = 0.001m$ in Algorithm~\ref{alg:mega}. Notice that the RPCA method is `semi-blind' that requires ${\bm Z}$; while the PCA method and Algorithm~\ref{alg:mega} are `fully blind' algorithms which do not require additional inputs other than the observed graph signals $\{ {\bm y}_\ell\}_{\ell=1}^m$ and the estimate of $k$. Lastly, in \Cref{alg:mega}, we estimate $\widehat{\bm Z}$ via Algorithm~\ref{alg:nmf}, which is initialized by setting each $\widehat{Z}_{ij}^{(0)}$, $\widehat{B}_{ij}^{(0)}$ to be ${\cal U}([0,1])$, and is terminated with a fixed number of iteration at $T=10^4$ to ensure convergence to optimal solution. For the subroutine \Cref{alg:nmf}, the stepsize parameters $a,b$ [cf.~\eqref{eq:stepsize}] are set as: ({\sf a}) $a=b=0.1$ for ${\cal H}_{\sf weak}({\bm A})$ with {\sf CP} graph, ({\sf b}) $a=b=0.01$ for ${\cal H}_{\sf strong}({\bm A})$ with {\sf CP} graph, ({\sf c}) $a=b=0.01$ with {\sf BA} graph.
{In addition, we compare with the natural heuristics which learns the graph topology from graph signals, and then detect the central nodes using the eigen-centrality vector computed from the estimated adjacency/Laplacian matrix. We benchmark with graph learning methods including {\sf GL-SigRep} \cite{dong2016learning}, {\sf SpecTemp} \cite{segarra2017network}, the method by Kalofolias \cite{kalofolias2016learn}, as well as the $k_{\sf NN}$-nearest neighbor ({\sf kNN}) graph constructed by setting an edge between a node and its $k_{\sf NN}=0.1n$ most correlated neighbors in the observed graph signals. 
}

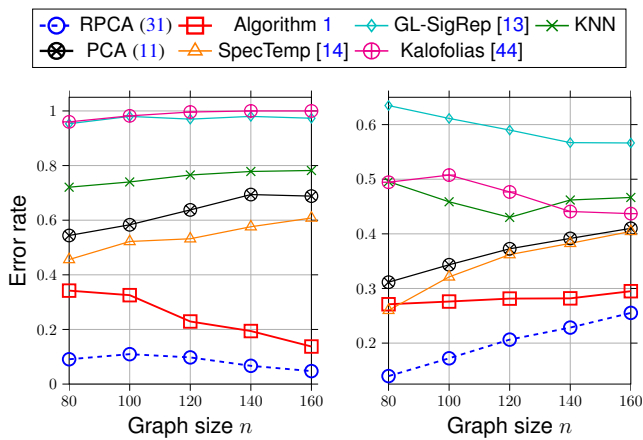
\begin{figure}[t]
\centering
{\sf 
\resizebox{.9\linewidth}{!}{\input{legend_tikz.tex}}}\\[.3cm]
{\sf 
\resizebox{!}{.53\linewidth}{\input{Evsn_wlp_cp}}
\resizebox{!}{.53\linewidth}{\input{Evsn_wlp_ba}}
}\vspace{-0.0cm}
\caption{Error rate of the central nodes detection methods against the size of graph $n$ for (Left) {\sf CP} and (Right) {\sf BA} with excitation's rank $k=50$ on ${\cal H}_{\sf weak}' ({\bm A})$. 
}\label{fig:Evsn}
\end{figure}

The first example focuses on the effect of excitation's rank, $k$, on the detection performance. The number of graph signal samples is $m=200$ and we consider {\sf CP} graphs with $p_1 = 0.4, p_2 = 0.05$ and $n=100$ nodes. 
In Fig.~\ref{fig:Evsk} (Left), we focus on the case with \emph{weak} low pass graph filter and compare the error rates in computing the top-10 central nodes of different algorithms against $k$. We observe that the error rate for the proposed methods generally decreases when $k$ increases. The semi-blind RPCA method obtains the best performance, followed by the fully blind \Cref{alg:mega}. Importantly, \Cref{alg:mega} delivers a significantly lower error rate compared to the methods which learn the complete graph topology. With a latent dimension of $k=40$, \Cref{alg:mega} detects the central nodes with an error rate of $\sim\!\!0.2$, which is three-fold lower than PCA and the other graph learning methods.
{Lastly, we observe an increased error rate for \Cref{alg:mega} at $k=50$. This is because as discussed in Section~\ref{sec:weak}, the NMF identifiability condition becomes more difficult to satisfy as the ratio $\frac{n}{k}$ decreases, see Fig.~\ref{fig:Evsn} for a further investigation on this phenomena.}
Furthermore, in Fig.~\ref{fig:Evsk} (Right), we consider the \emph{strong} low pass graph filter. The proposed methods obtain almost zero error rates over the  range of $k$ tested.

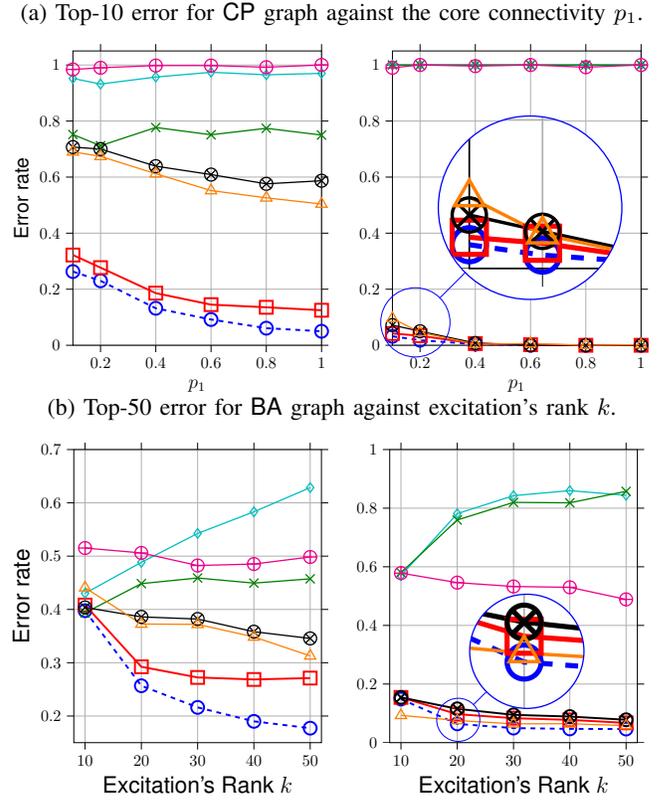
\begin{figure}[t]
\centering
{\sf 
\begin{subfigure}{1\linewidth}
\centering
\caption{Top-10 error for {\sf CP} graph against the core connectivity $p_1$.}
\resizebox{!}{.53\linewidth}{\input{Evsp_wlp_cp}}
\resizebox{!}{.53\linewidth}{\input{Evsp_slp_cp}}
\end{subfigure}
\begin{subfigure}{1\linewidth}
\centering
\caption{Top-50 error for {\sf BA} graph against excitation's rank $k$.}{
\resizebox{!}{.54\linewidth}{\input{Evsk_wkp_ba}}
\resizebox{!}{.54\linewidth}{\input{Evsk_slp_ba}}
} 
\end{subfigure}}
\caption{Error rate comparison for the central nodes detection methods. (Left) Weak low pass filter ${\cal H}_{\sf weak}({\bm A})$. (Right) Strong low pass filter ${\cal H}_{\sf strong}({\bm A})$. Readers are referred to Fig.~\ref{fig:Evsk} for legend of the above plots.}\label{fig:Evsm}\vspace{-.2cm}
\end{figure}

{The second example considers the effect of the core-periphery structure on the detection performance. Specifically, in Fig.~\ref{fig:Evsm} (a), we focus on the {\sf CP} graph with $n=100$ nodes, $p_2 = 0.05$, and compare the error rates against the connectivity parameter among the core nodes, $p_1 \in \{0.1,...,1.0\}$. The excitation's rank is fixed at $k=40$. As observed, the error rates generally decrease as $p_1$ increases for both strong and weak low pass graph filters. This is anticipated since the spectral gap in the adjacency matrix $\frac{\lambda_2}{\lambda_1}$ improves as the graph exhibits a stronger core-periphery structure. Moreover, the proposed \Cref{alg:mega} and RPCA method \eqref{eq:rpca} outperform the benchmark algorithms.

The third example tests the performance of detecting central nodes when the underlying graph is the {\sf BA} graph with $n=100$ nodes. Notice that we focus on the error rate in detecting the top-$C=50$ central nodes defined by ${\bm c}_{\sf eig}$ of the adjacency matrix. We compare the error rate against the excitation's rank $k$. The result is shown in Fig.~\ref{fig:Evsm} (b). As in the first example, for both strong and weak low pass graph filters, we observe that the error rate gradually decreases as $k$ increases, and the proposed \Cref{alg:mega} outperforms the benchmark algorithms. However, we note that the performance gap is not as significant as in the case of {\sf CP} graphs. A possible reason for this is that the {\sf BA} graphs have a weaker core-periphery structure than {\sf CP} graphs; in fact, nodes with similar magnitude in the eigen-centrality vector ${\bm c}_{\sf eig}$ may be misidentified.}

\begin{figure}[t]
\begin{center}
    \includegraphics[width=.48\linewidth]{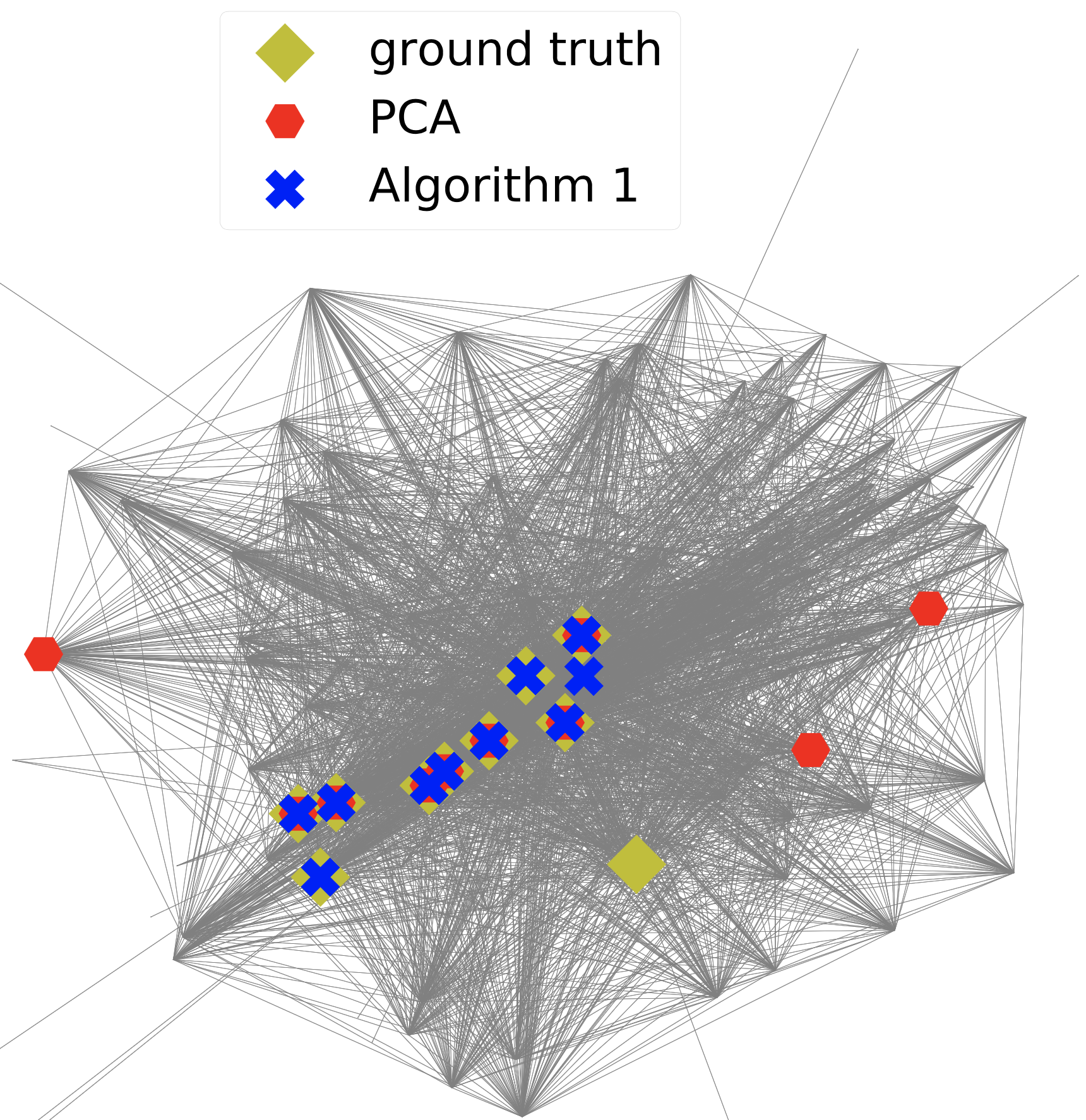}~
    \includegraphics[width=.5\linewidth]{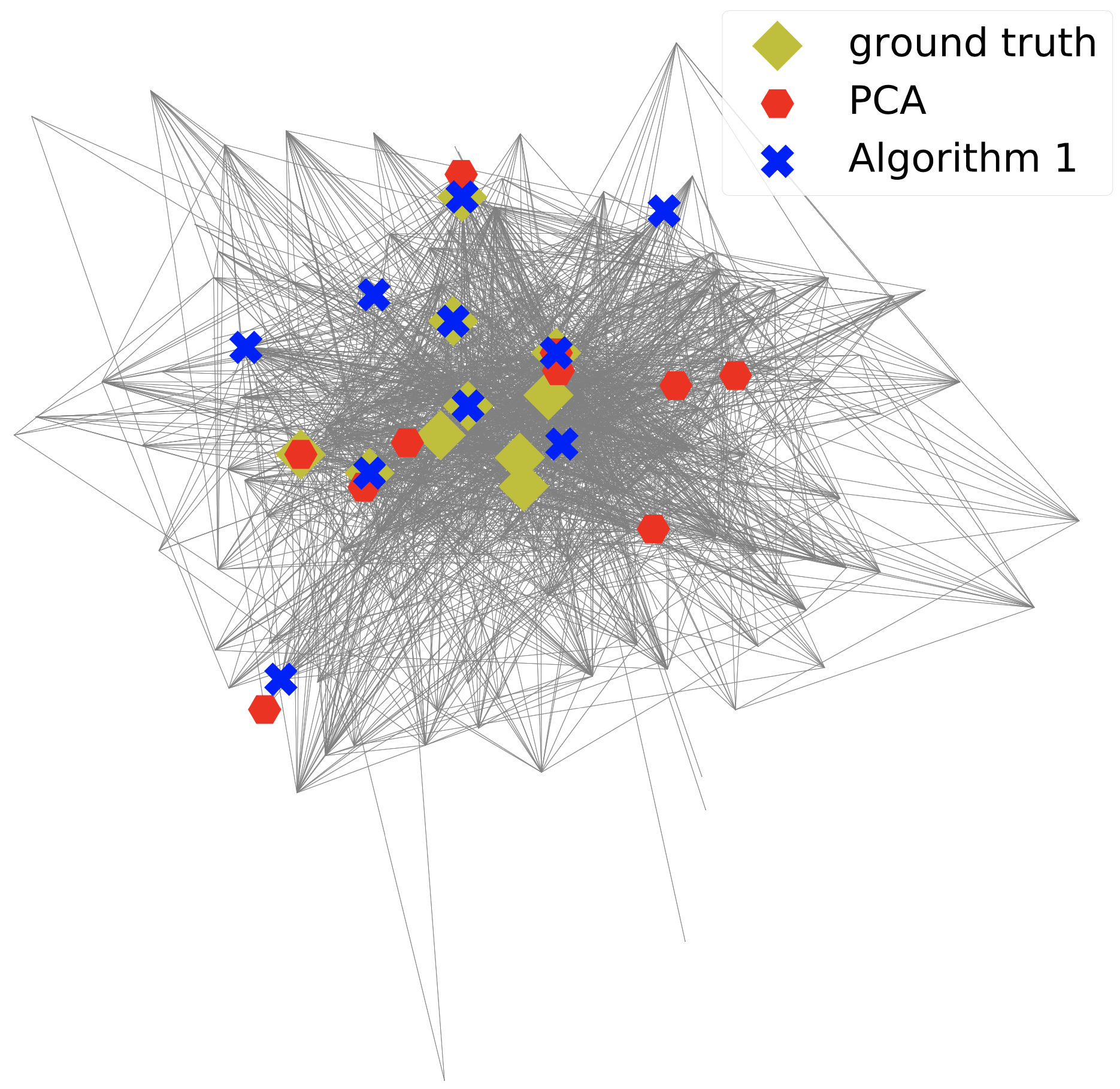}\vspace{-.0cm}
\end{center}
\caption{Example illustrating the detection results of the proposed methods: (Left) {\sf Email} graph and (Right) {\sf Neural} graph with weak low pass filter ${\cal H}_{\sf weak}'' ({\bm A})$. Step sizes parameters \eqref{eq:stepsize} are set as  $a=b=0.01$.}\label{fig:realfilterreliaz}\vspace{-.2cm}
\end{figure}

{The fourth example compares the detection performances with respect to the graph size $n$. We focus on the weak low pass graph filter ${\cal H}_{\sf weak}' ({\bm A}) = ( {\bm I} - \frac{2}{ n } {\bm A} )^{-1}$ and fix the excitation's rank at $k = 50$. The results are shown in Fig.~\ref{fig:Evsn} for the {\sf CP} (with $p_1 = 0.4$) and {\sf BA} graph models as we evaluate the top-10 and top-$0.5n$ error rates, respectively. Notice that for \Cref{alg:mega}, the NMF stage favors models with large $\frac{n}{k}$ as it leads to easier satisfaction of the identifiabiltiy conditions in \Cref{fact:unique}. The figure shows that the error rate of \Cref{alg:mega} approaches that of the {\sf RPCA} as $n$ increases, as predicted by our analysis. Notice that the performances of all algorithms deteriorate with the graph size $n$ for {\sf BA} graph as the eigen-centrality vector of {\sf BA} graphs is less localized.

We also compare the detection results of the proposed methods by simulating graph signals on real graph topologies. We consider the {\sf Email} and {\sf Neural} graphs taken from the \texttt{KONECT} project (available: \url{http://konect.cc}). The {\sf Email} graph has $n=167$ nodes and the {\sf Neural} graph has $n=297$ nodes. We generate the graph signals with a weak low pass graph filter ${\cal H}_{\sf weak}'' ({\bm A}) = ( {\bm I} - \frac{1}{ 100 } {\bm A} )^{-1}$ and the excitation's rank is fixed at $k=\lceil 0.3n \rceil$. The graphs together with the detection results of our algorithms are presented in Fig.~\ref{fig:realfilterreliaz}. The {\sf Email} graph has a core-periphery structure where $\frac{\lambda_2}{\lambda_1} = 0.254$, while the {\sf Neural} graph does not have a set of significantly core nodes as  $\frac{\lambda_2}{\lambda_1} = 0.585$. We observe that \Cref{alg:mega} performs well on the {\sf Email} graph; while the central nodes detection are not as accurate on the {\sf Neural} graph. 
}

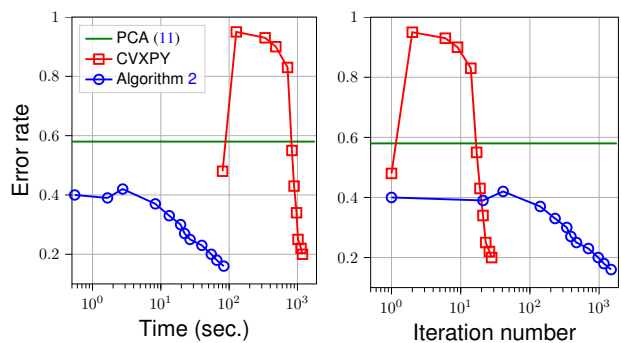
\begin{figure}[t]
\centering
{\sf \resizebox{!}{.51\linewidth}{\input{Evst_time}}\resizebox{!}{.51\linewidth}{\input{Evst_iterv2}}}\vspace{-0.0cm}
\caption{Error rate of central nodes detection implementations with \Cref{alg:mega} the (Left) running time and (Right) iteration number. 
} \label{fig:Evstime}\vspace{-.3cm}
\end{figure}

As the last example, we examine the efficient implementation of \Cref{alg:mega}.
In Fig.~\ref{fig:Evstime}, we fix $n = 100, m =200$, $k = 40$, $a=b=0.1$, consider the {\sf CP} graph with $p_1 = 1, p_2 = 0.05$, and the graph filter ${\cal H}_{\sf weak}({\bm A}) = ( {\bm I} - \frac{1}{50} {\bm A} )^{-1}$. We compare the error rates against running time and iteration for \Cref{alg:nmf} and the \texttt{cvxpy}-based algorithm. 
Both algorithms converge to a solution with similar error rates. 
In particular, the \texttt{cvxpy}-based algorithm  converges in less than 60 iterations, while \Cref{alg:nmf} requires more than $1000$ iterations to converge. 
However, in terms of the running time, \Cref{alg:nmf} is around 10 fold faster than \texttt{cvxpy}-based algorithm.\vspace{-.2cm}

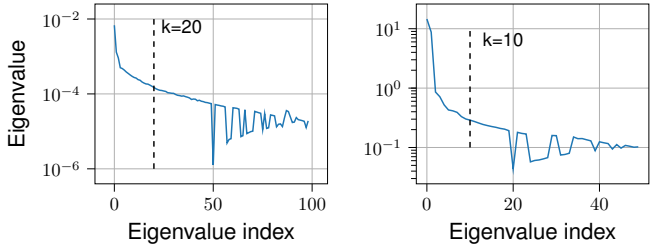
\begin{figure}[t]
\centering
{\sf 
\resizebox{!}{.37\linewidth}{\input{Eigenvalue_stock}}
\resizebox{!}{.37\linewidth}{\input{Eigenvalue_vote}}
}\vspace{-0.0cm}
\caption{Spectrum of the data covariance matrix. (Left) {\sf Stock} dataset and (Right) {\sf Senate} dataset. {\blue Note that the ratio between the second and first eigenvalue of the covariance matrix is 0.19 for {\sf Stock} dataset and 0.60 for {\sf Senate} dataset.}}\vspace{-0.2cm}\label{fig:data}
\end{figure}

\begin{table*}[t]
    \setlength{\tabcolsep}{2.5pt}
    \centering
    \scriptsize
    \begin{subtable}[t]{.575\linewidth}
    \caption{{\sf Stock} Dataset$^\dagger$}
    \resizebox{0.975\width}{!}{\begin{tabular}{l l l l l l l l l l l}
    \toprule
    \bfseries Method & \multicolumn{10}{c}{\bfseries Top-10 Estimated Central Stocks (sorted left-to-right)}\\ \midrule
     \Cref{alg:mega} & {\orange ALL} & {\red ACN} & {\green HON}  & {\orange AXP} & {\red IBM} & {\color{blue} DIS} & {\red ORCL} & {\green MMM} & {\orange BRK.B} & COST \\
     & 0.43 & 0.56 & 0.51 & 0.72 & 0.50 & 0.36 & 0.70 & 0.33 &  0.52 & 0.64\\
    & \multicolumn{10}{c}{Average Correlation Score: $0.53 \pm 0.133$}\\
    \hline
    \rowcolor{gray!10} \textbf{PCA} \eqref{eq:pca} & {\red NVDA} & {\color{blue} NFLX} & AMZN & {\red ADBE} & {\orange PYPL} & {\green CAT} & {\orange MA} & {\color{blue} GOOG} & {\green BA} & {\color{blue} GOOGL}  \\
    \rowcolor{gray!10} 
    & 0.56 & 0.60 & 0.68 & 0.63 & 0.65 & 0.27 & 0.67 &  0.63 & 0.28 & 0.63\\
    \rowcolor{gray!10}& \multicolumn{10}{c}{\textbf{Average Correlation Score:} ${\bf 0.56 \pm 0.154}$}\\
    \hline 
    {\sf GL-SigRep} & {\color{blue} GOOGL} & {\color{blue} GOOG} & LLY & {\orange USB} & {\green EMR} & DUK & {\red ORCL} & {\green GD} & {\color{blue} VZ} & {\orange V}\\
    \cite{dong2016learning}
    & 0.63 & 0.63 & 0.17 &0.43 &0.59 & 0.11 &  0.70 &0.53 &0.27 & 0.71\\
    & \multicolumn{10}{c}{Average Correlation Score: $0.48 \pm 0.22$}\\
    \hline
    KNN & {\red ACN} & {\green HON}  & {\orange ALL} & {\orange BRK.B} & {\red IBM} & {\orange AXP} & {\green EMR}  & {\green MMM} & {\red CSCO} & XOM\\
    & 0.56 & 0.51 & 0.43 & 0.52 & 0.50 & 0.72 & 0.59 &0.33 & 0.63 & 0.55\\
    & \multicolumn{10}{c}{Average Correlation Score: $0.53 \pm 0.107$}\\
    \hline
    {\sf SpecTemp} & {\red ACN} & {\red ORCL} & PG & LLY & SUBX & {\orange PYPL} & MDLZ & {\color{blue} FB} & PFE & MRK\\
    \cite{segarra2017network}
    & 0.56 & 0.70 & 0.36 & 0.17 & 0.58 & 0.65 & 0.41 & 0.61 & 0.14 & 0.20\\
    & \multicolumn{10}{c}{Average Correlation Score: $0.44 \pm 0.211$}\\
    \hline
    Kalofolias & {\red ACN} & {\green HON} & {\orange BRK.B}   & {\orange ALL} & {\orange AXP} & {\red IBM} & XOM & KO& {\orange USB} & COST   \\
    \cite{kalofolias2016learn}
    & 0.56 & 0.51 & 0.52 & 0.43 & 0.72 & 0.50 & 0.55 & 0.32 & 0.43 & 0.64\\
    & \multicolumn{10}{c}{Average Correlation Score: $0.52 \pm 0.112$}\\
    \hline 
     \multicolumn{11}{c}{{\red Information Technology}/ {\color{blue} Communication Services}/ {\green Industrials}/ {\orange Financials}/other sectors. }\\
    \bottomrule
    \end{tabular}}
    \end{subtable}
    \begin{subtable}[t]{.42\linewidth}
    \caption{{\sf Senate} Dataset$^\dagger$}
    \resizebox{0.975\width}{!}{\begin{tabular}{l l l l l l l l l l l}
    \toprule
    \bfseries Method & \multicolumn{10}{c}{\bfseries Top-10 Estimated Central States (sorted left-to-right)}\\ \midrule
    \rowcolor{gray!10}
    \rowcolor{gray!10} \textbf{\Cref{alg:mega}} & {\color{blue} MI} & {\color{blue} MT} & {\red KS} & {\color{blue} RI} & {\red TN} & {\green MN} & {\green NV} & {\red ME} & {\color{blue} MD} & {\green IN} \\ 
    \rowcolor{gray!10} 
    & 0.79 & 0.66 &  0.74 & 0.67 & 0.68 & 0.74 & 0.43 & 0.67 & 0.6 & 0.62\\
    \rowcolor{gray!10} & \multicolumn{10}{c}{ \textbf{Average Correlation Score:} ${\bf 0.66 \pm 0.099}$}\\
    \hline
    PCA \eqref{eq:pca} & {\color{blue} CA} & {\color{blue} DE} & {\green CO} & {\color{blue} IL} & {\color{blue} ND} & {\color{blue} WV} & {\green IA} & {\green VA} & {\red WY} & {\color{blue} MA}\\
    & 0.55 & 0.46 & 0.54 & 0.63 &  0.72 & 0.52 & 0.51 & 0.56 & 0.59 & 0.58\\
    & \multicolumn{10}{c}{Average Correlation Score: $0.57 \pm 0.072$}\\
    \hline 
    {\sf GL-SigRep} & {\color{blue} CA} & {\color{blue} DE} & {\color{blue} WV} &{\green CO} & {\color{blue} IL} & {\green VA} & {\color{blue} ND} & {\green IA} & {\red WY} & {\red AZ}\\
    \cite{dong2016learning}
    & 0.55 & 0.46 & 0.52 & 0.54 & 0.63 & 0.56 & 0.72 & 0.51 & 0.59 & 0.31\\
    & \multicolumn{10}{c}{Average Correlation Score: $0.54 \pm 0.108$}\\
    \hline
    KNN & {\color{blue} ND} & {\color{blue} CA} & {\color{blue} IL} & {\color{blue} WV} & {\color{blue} DE} & {\green VA} & {\red AZ} & {\green CO} & {\red WY} & {\green IA}\\
    & 0.72 & 0.55 & 0.63 & 0.52 & 0.46 & 0.56 & 0.31 & 0.54 & 0.59 & 0.51\\
    & \multicolumn{10}{c}{Average Correlation Score: $0.54 \pm 0.108$}\\
    \hline
    {\sf SpecTemp} & {\red AL} & {\color{blue} ND} & {\color{blue} WV} & {\color{blue} CA} & {\color{blue} DE} & {\color{blue} IL} & {\green MO} & {\color{blue} MA} & {\green VA} & {\green SD}\\
    \cite{segarra2017network}
    & 0.61 &  0.72 & 0.52 & 0.55 & 0.46 & 0.63 & 0.57 &  0.58 & 0.56 & 0.56\\
    & \multicolumn{10}{c}{Average Correlation Score: $0.58 \pm 0.069$}\\
    \hline
    Kalofolias &{\red AL} & {\red AK} & {\red AZ} & {\color{blue} AR} & {\color{blue} WV} & {\green VA} & {\color{blue} CA} & {\green CO} & {\color{blue} CT} & {\color{blue} DE}\\
    \cite{kalofolias2016learn} 
    & 0.61 & 0.63 & 0.31 & 0.47 & 0.52 & 0.56 & 0.55 & 0.54 & 0.45 & 0.46\\
    & \multicolumn{10}{c}{Average Correlation Score: $0.51 \pm 0.093$}\\
    \hline 
     \multicolumn{11}{c}{{\red Republican}/ {\color{blue} Democrat}/ {\green Mixed}.}\\
    \bottomrule
    \end{tabular}}\vspace{.3cm}
    \end{subtable}
    \footnotesize{$^\dagger$The number below each stock/state shows its normalized correlation score with the S\&P100 index and number of `Yay's in the voting result [cf.~\eqref{eq:corr}]. The average correlation scores are taken over the set of central nodes found and the number after `$\pm$' is the standard deviation. 
    }
    \caption{Estimated Central Stocks/States from the {\sf Stocks}/{\sf Senate} Datasets.}\vspace{.0cm}
    \label{table:top 10} 
\end{table*}

\subsection{Experiments on Real Data}\vspace{-.1cm}
In this subsection, we experiment on detecting the central nodes of the latent graph from two datasets of graph signals. The first dataset ({\sf Stock}) is the daily return from S\&P100 stocks in May 2018 to Aug 2019 with $n=99$ stocks, $m=300$ samples from \url{https://www.alphavantage.co/}\footnote{{\blue The {\sf Stock} dataset is pre-processed by subtracting the daily returns by the minimum return value across all samples.}}. The second dataset ({\sf Senate}) contains $m = 657$ votes grouped by $n=50$ states at the US Senate in 2007 to 2009 from \url{https://voteview.com}, and consider the combined votes from 2 Senators of the same state by assigning a score of $+2, +1, 0$ for a `Yay', `Abstention', `Nay' vote, respectively. 
The spectrum of the two datasets' covariance matrices are plotted in Fig.~\ref{fig:data}. For {\sf Stock} dataset, the excitation's rank is estimated at $k = 20$; for {\sf Senate} dataset, the excitation's rank is estimated at $k = 10$.
To remove bias from random initialization for \Cref{alg:mega}, we run the algorithm for 100 times and record the frequencies in which a node is ranked among the top-$C$ in the estimated centrality ranking. The nodes with the highest $C$ frequencies are denoted as the estimated top-$C$ central nodes. 

{\blue As the actual central nodes are unknown, we consider testing the quality of selected central nodes as \emph{predictors} for the network's overall outcome such as the S\&P100 index's daily return ({\sf Stock} dataset) or voting results ({\sf Senate} dataset). To setup the experiment, we take the first $80\%$ of samples as training data and the last 20\% as testing data. We apply \Cref{alg:mega}, PCA and other benchmark methods to estimate the top-$10$ central nodes from the \emph{training data}. Then, for each central node, we compute the \emph{normalized correlation score} between the time series of that node over the \emph{testing data} and the overall outcome over the matching days/rollcall. For node $i$, we compute
\beq \label{eq:corr}
{\sf corr}_i = \| [ {\bm Y} ]_{i, {\sf Test}} \|^{-1} \| {\bm g} \|^{-1} \langle [ {\bm Y} ]_{i, {\sf Test}} , {\bm g} \rangle,
\eeq 
where ${\bm g}$ is the S\&P100 index's returns (for {\sf Stock} dataset), or the number of `Yay' votes, for the days or rollcalls (for {\sf Senate} dataset) corresponding to \emph{testing data}. Notice that ${\sf corr}_i \in [0,1]$ as the signals are non-negative. A higher ${\sf corr}_i$ indicates that node $i$ is a good predictor for the global behavior of all nodes, which may correspond to a more central node. 
}


Table~\ref{table:top 10}-(a) shows the estimated top-$C=10$ central nodes for {\sf Stock} dataset. We observe that PCA identifies the technology firms (e.g., NVDA, NFLX, AMZN); while \Cref{alg:mega} identifies a diverse list of firms, ranging from {\blue finance (e.g. ALL), industry (e.g. HON), to technology (e.g. ACN, IBM) firms.}
{\blue Among the benchmarked methods, PCA achieves the highest average correlation scores, and \Cref{alg:mega} has a similar performance. This can be explained by the large gap between the first and second eigenvalues in this dataset (see Fig.~\ref{fig:data} (left)), which suggests that the graph signals may have been filtered by a \emph{strong} low pass filter.}

Table \ref{table:top 10}-(b) considers the {\sf Senate} dataset and highlights the top-$C=10$ states identified. Compared to the other benchmarked methods, we first observe that \Cref{alg:mega} identifies a more `balanced' set of states with sitting Senators from different parties in 2007-2009. {\blue Moreover, \Cref{alg:mega} finds a set of states with a significantly higher average correlation score than the other algorithms. Notice that from Fig.~\ref{fig:data} (right), this dataset has a small gap between the first and second eigenvalues, suggesting that the underlying low pass graph filter maybe \emph{weak}. In the latter scenario, \Cref{alg:mega} can reliably detect the central nodes. }

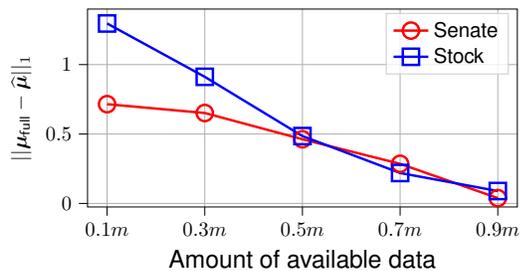
\begin{figure}[t]
\centering
{\sf 
\resizebox{!}{.4\linewidth}{\input{consist}}
}\vspace{-0.0cm}
\caption{Distance in distribution on real datasets against the amount of available data.}\vspace{-0.4cm}\label{fig:consistency}
\end{figure}

{\blue Lastly, we examine the consistency of \Cref{alg:mega} on both datasets. We select a subset of samples randomly from the dataset with sizes $0.1m, 0.3m, \ldots, 0.9m$. For each subset, we record the frequencies in which a node is chosen as one of the top-10 central nodes using 100 trials with random initialization to \Cref{alg:mega}, and construct the empirical distribution $\widehat{\bm \mu}$ accordingly. 
We then compare $\widehat{\bm \mu}$ to the empirical distributed constructed similarly by applying \Cref{alg:mega} to the full dataset, the latter is denoted as ${\bm \mu}_{\sf full}$.
Fig.~\ref{fig:consistency} compares the distance in distribution as $\| \widehat{\bm \mu} - {\bm \mu}_{\sf full} \|_1$ averaged over 100 randomly selected subsets. As the amount of available data increases, the distance in distribution decreases as \Cref{alg:mega} produces more consistent estimates of the central nodes. Notice that for {\sf Stock} dataset, the larger distance in distribution reflects the higher volatility of the stock return data.
}\vspace{-.2cm}

\section{Conclusions}\vspace{-.2cm}
{We have studied the problem of blind central nodes detection under a variety of settings. Our method only relies on a mild condition that the underlying graph filter is low pass, while the excitation can be low-rank in general. 
For strong low pass graph filter, we show that the PCA method detects the central nodes correctly. 
For general low pass graph filter, we study a structured factor analysis model and analyse its identifiability through treating a sparse NMF model. The latter motivated us to develop a two-stage decomposition algorithm combining NMF and RPCA. In all settings, we analyze the problem parameters affecting the estimation performance. 
Future works include incorporating advanced NMF criterion from \cite{fu2019nonnegative} into \Cref{alg:mega}, and analysis for the recoverability of the studied structured factor model with noise.\vspace{-.2cm}}

\appendices
\section{Proof of Lemma~\ref{lem:perfbd}}\label{sec:lemma1}\vspace{-.1cm}
Since $||\bar{\bm v}_1|| = 1$, $||{\bm c}_{\sf eig}|| =1$, we have $(\bar{\bm v}^\top_1{\bm c}_{\sf eig})^2 \leq \bar{\bm v}^\top_1{\bm c}_{\sf eig}$ and it holds that
\beq \notag
\begin{split}
& ||\widehat{\bm v}_1-{\bm c}_{\sf eig}||^2 = ||\widehat{\bm v}_1||^2 +||{\bm c}_{\sf eig}||^2 -2 \cdot \widehat{\bm v}^\top_1{\bm c}_{\sf eig}\\
&\leq || \widehat{\bm v}_1||^2+||{\bm c}_{\sf eig}||^2-2 \cdot (\widehat{\bm v}^\top_1{\bm c}_{\sf eig})^2 = || \widehat{\bm v}_1 \widehat{\bm v}_1^\top-{\bm c}_{\sf eig}{\bm c}_{\sf eig}^\top||^2_F\\
&\leq 2 \cdot || \widehat{\bm v}_1 \widehat{\bm v}^\top_1 -{\bm c}_{\sf eig}{\bm c}_{\sf eig}^\top||^2.
\end{split}
\eeq 
This implies $||\widehat{\bm v}_1-{\bm c}_{\sf eig}|| \leq \sqrt{2} || \widehat{\bm v}_1 \widehat{\bm v}^\top_1 -{\bm c}_{\sf eig}{\bm c}_{\sf eig}^\top||$. Note that the latter is the spectral norm of a rank-2 matrix.
Furthermore, the triangular inequality yields:
\beq\label{eq:prop_split}
\begin{split}
&||\widehat{\bm v}_1\widehat{\bm v}_1^{\top} - {\bm c}_{\sf eig} {\bm c}_{\sf eig}^\top || \\
&\leq ||\bar{\bm v}_1\bar{\bm v}_1^{\top} - {\bm c}_{\sf eig} {\bm c}_{\sf eig}^\top || + ||\widehat{\bm v}_1\widehat{\bm v}_1^{\top}-\bar{\bm v}_1\bar{\bm v}_1^{\top}||.
\end{split}
\eeq
The first term is bounded by \cite[Proposition 1]{wai2019blind} with $K=1$:
\beq \label{eq:prop_bd}
\| \bar{\bm v}_1 \bar{\bm v}_1^\top - {\bm c}_{\sf eig} {\bm c}_{\sf eig}^\top \|
= \sqrt{ \gamma^2 / (1+\gamma^2) } \leq \gamma,
\eeq
with $\gamma \leq \eta \frac{ \| {\bm V}_{N-1}^\top {\bm B} {\bm q}_1 \| }{ |{\bm v}_1^\top {\bm B} {\bm q}_1| }$. 
The second term is bounded by \cite[Proposition 2]{wai2019blind} with $K=1$:
\beq \label{eq:prop_bd2}
||\widehat{\bm v}_1\widehat{\bm v}_1^{\top}-\bar{\bm v}_1\bar{\bm v}_1^{\top}|| \leq \delta^{-1} {||\bm{\Delta}||}.
\eeq 
Combining the above inequalities yield the desirable results.\vspace{-.2cm}

\section{Proof of Lemma~\ref{lem:sparse}} \label{app:sparse}
For any $\rho > 0$, we observe that 
\beq \label{eq:bound_needed}
\frac{ \| \tilde{\cal H}_\rho ({\bm A}) {\bm B} \| }{ \rho \| {\bm B} \| } \leq \frac{ \| \tilde{\cal H}_\rho ({\bm A}) \| }{ \rho }.
\eeq
Taking $\min_{ \rho > 0 }$ on both sides yields an upper bound to the ratio in \eqref{eq:sparse_ratio} as
\beq \label{eq:min_rho}
\min_{ \rho > 0 } \!~ f(\rho) := \max \{ | \rho^{-1} h(\lambda_1)  - 1|, ..., | \rho^{-1} h(\lambda_n)  - 1| \}.
\eeq
Under Assumption~\ref{assump:lowpass}, we note that $h_{\sf min} < h(\lambda_1)$ and there are at least two distinct values in the collection $\{ h(\lambda_1),...,h(\lambda_n)\}$. Furthermore, for $0 \in \partial f(\rho^\star)$, we must have $\rho^\star$ satisfying 
\beq
| (\rho^\star)^{-1} h(\lambda_1) - 1 | = | (\rho^\star)^{-1} h_{\sf min} - 1 | .
\eeq
Since $h(\lambda_1) \neq h_{\sf min}$, it holds
\beq
\rho^\star = 2^{-1} ( h(\lambda_1) + h_{\sf min} ).
\eeq
{Evaluating $f( \rho^\star )$ and simplifying the expression gives the bound
\beq \label{eq:sparse_ratio_pre}
\min_{ \rho > 0 } \frac{ \| \tilde{\cal H}_\rho ({\bm A}) {\bm B} \| }{ \rho \| {\bm B} \| } \leq \frac{ h(\lambda_1) - h_{\sf min} }{ h(\lambda_1) + h_{\sf min} }.
\eeq
As the frequency response $h(\lambda)$ is a convex and non-negative function, it holds that $\eta = \max\{ h(\lambda_2), h(\lambda_n) \} / h(\lambda_1)$. Therefore, dividing the nominator and denominator in the r.h.s.~of \eqref{eq:sparse_ratio_pre} by $h(\lambda_1)$ yields the final bound in \eqref{eq:sparse_ratio}.\vspace{-.2cm}
}

\section{Proof of Lemma \ref{cor:diagonal}}\label{app:diagonal}
Applying \Cref{cor:perf} to $\widehat{\bm L}$ in \eqref{eq:rpca_approx} yields:
\beq \label{eq:app_upbd}
\| \widetilde{\bm v}_1 - {\bm c}_{\sf eig} \| \leq \widetilde{\eta} \cdot \sqrt{2} \frac{ \| {\bm V}_{n-1}^\top {\bm B} {\bm D} \bm{\Pi} {\bm q}_1 \|_2 }{ |{\bm v}_1^\top {\bm B} {\bm D} \bm{\Pi} {\bm q}_1| }.
\eeq
Since ${\rm diag}({\bm D}) > {\bm 0}$, we observe that
\beq \label{eq:firstbd_n}
\frac{ \| {\bm V}_{n-1}^\top {\bm B} {\bm D} \bm{\Pi} {\bm q}_1 \| }{ |{\bm v}_1^\top {\bm B} {\bm D} \bm{\Pi} {\bm q}_1| } 
\leq \frac{ \| {\bm V}_{n-1}^\top {\bm B} \bm{\Pi} {\bm q}_1 \| }{ |{\bm v}_1^\top {\bm B} \bm{\Pi} {\bm q}_1| } \frac{ \max_{i} D_{i,i} }{ \min_i D_{i,i} }.
\eeq
As we have taken $\widehat{\bm Z}{\bf 1} = {\bm 1}$, it holds
\beq
{\bf 1} = \widehat{\bm Z} {\bf 1} = \bm{\Pi}^\top {\bm D}^{-1} {\bm Z} {\bf 1} \Longleftrightarrow {\rm diag}({\bm D}) = {\bm Z} {\bf 1}.
\eeq
It shows that the diagonal scaling of the obtained solution is controlled by the random vector ${\bm Z} {\bf 1}$. Under \Cref{ass:const}, for any $t > 0$, the Hoeffding's inequality implies that 
\beq \textstyle 
\PP( \| {\bm Z} {\bf 1}/m - \alpha {\bf 1} \|_\infty \geq t ) \leq 2k \exp( - mt^2 / (2 \sigma_z^2) ).
\eeq
Therefore, with probability at least $1-2k \exp( - mt^2 / (2 \sigma_z^2) )$, 
\beq
\frac{ \max_{i} D_{i,i} }{ \min_i D_{i,i} } = \frac{ \max_i \sum_{\ell=1}^m z_{\ell,i}/m }{ \min_i \sum_{\ell=1}^m z_{\ell,i}/m } \leq \frac{\alpha+t}{\alpha-t}.
\eeq 
{Setting $\exp( - mt^2 / (2 \sigma_z^2) ) = \delta$ and substituting the above into \eqref{eq:firstbd_n} conclude the proof.\vspace{-.2cm}}

\section{Proof of Fact~\ref{fact:conv}} \label{app:factconv}
Given an initialization $\hat{\bm B}^{(0)}, \hat{\bm Z}^{(0)}$, define the following compact subset of $\Re^{n \times k}$:
\beq \textstyle
{\cal B}^0 = \{ {\bm B}  : {\bm B} \geq {\bm 0}, \sum_{ij} B_{ij} \leq \lambda_S^{-1} f( \hat{\bm B}^{(0)}, \hat{\bm Z}^{(0)} ) \}.
\eeq
With our choice of step sizes in \eqref{eq:stepsize}, Algorithm~\ref{alg:nmf} returns iterates which satisfy $\hat{\bm B}^{(t)} \geq {\bm 0}$ and
\beq
f( \hat{\bm B}^{(t+1)}, \hat{\bm Z}^{(t+1)} ) \leq f( \hat{\bm B}^{(t)}, \hat{\bm Z}^{(t)} ),~\forall~t \geq 0,
\eeq
see \cite{wu2020hybrid}. 
As the squared Frobenius norm $\frac{1}{2}\| {\bm Y} - {\bm B} {\bm Z} \|_F^2$ is non-negative, the above implies that
\beq
\textstyle
\lambda_S \sum_{ij} \hat{B}_{ij}^{(t)} \leq f( \hat{\bm B}^{(0)}, \hat{\bm Z}^{(0)} ),
\eeq
and thus $\hat{\bm B}^{(t)} \in {\cal B}^0$ for all $t \geq 0$.

We then observe that Algorithm~\ref{alg:nmf} is a special case of the hybrid BCD algorithm applied to:
\beq
\min_{ {\bm B}, {\bm Z} } \widetilde{f}( {\bm B}, {\bm Z} ) \eqdef f( {\bm B}, {\bm Z} ) + {\cal I}_{ {\cal B}^0 } ( {\bm B} ) + {\cal I}_{\Delta^K} ( {\bm Z} ),
\eeq
where ${\cal I}_{ {\cal B}^0 } (\cdot), {\cal I}_{\Delta^K} (\cdot)$ are the $0/\infty$ indicator functions of the sets ${\cal B}^0, \Delta^K$, respectively, and such that a PGD update step is selected for the individual block updates. Furthermore, the domain of $\widetilde{f}$ is a compact set ${\cal B}^0 \times \Delta^K$.
Consequently, the conclusion in the fact follows by applying \cite[Theorem 1]{wu2020hybrid}.\vspace{-.2cm}

\bibliographystyle{IEEEtran}
\bibliography{ref_list}

\end{document}

%% file: stopband.tex
\begin{tikzpicture}

\begin{axis}[
tick pos=left,
xmajorgrids,
ymajorgrids,
tick pos=left,
x dir=reverse,
xmin=-2, xmax=10,
ymin=0.0449924513497903, ymax=1.16187866633784,
xtick={8.98989898989899,1.11111111111111,-0.888888888888889},
xticklabels={$\lambda_1$,$\lambda_2$,$\dots\lambda_n$},
ytick={1.09877913429523,1.01123595505618},
yticklabels={\small $h_{\sf w}(\lambda_1)$, \small $h_{\sf w}(\lambda_2)$},
width=6cm,height=5cm,
axis y line=left,
]

\addplot [ultra thick, green!50.0!black,name path=A]
table {%
10 1.11111111111111
9.8989898989899 1.10986547085202
9.7979797979798 1.10862262038074
9.6969696969697 1.10738255033557
9.5959595959596 1.10614525139665
9.49494949494949 1.10491071428571
9.39393939393939 1.10367892976589
9.29292929292929 1.10244988864143
9.19191919191919 1.10122358175751
9.09090909090909 1.1
8.98989898989899 1.09877913429523
8.88888888888889 1.09756097560976
8.78787878787879 1.09634551495017
8.68686868686869 1.09513274336283
8.58585858585859 1.0939226519337
8.48484848484848 1.09271523178808
8.38383838383838 1.09151047409041
8.28282828282828 1.09030837004405
8.18181818181818 1.08910891089109
8.08080808080808 1.08791208791209
7.97979797979798 1.08671789242591
7.87878787878788 1.08552631578947
7.77777777777778 1.08433734939759
7.67676767676768 1.08315098468271
7.57575757575758 1.08196721311475
7.47474747474747 1.08078602620087
7.37373737373737 1.07960741548528
7.27272727272727 1.07843137254902
7.17171717171717 1.07725788900979
7.07070707070707 1.07608695652174
6.96969696969697 1.07491856677524
6.86868686868687 1.07375271149675
6.76767676767677 1.07258938244854
6.66666666666667 1.07142857142857
6.56565656565657 1.07027027027027
6.46464646464646 1.06911447084233
6.36363636363636 1.06796116504854
6.26262626262626 1.06681034482759
6.16161616161616 1.06566200215285
6.06060606060606 1.06451612903226
5.95959595959596 1.06337271750806
5.85858585858586 1.06223175965665
5.75757575757576 1.06109324758842
5.65656565656566 1.05995717344754
5.55555555555556 1.05882352941176
5.45454545454546 1.05769230769231
5.35353535353535 1.05656350053362
5.25252525252525 1.05543710021322
5.15151515151515 1.05431309904153
5.05050505050505 1.0531914893617
4.94949494949495 1.05207226354942
4.84848484848485 1.05095541401274
4.74747474747475 1.04984093319194
4.64646464646465 1.04872881355932
4.54545454545455 1.04761904761905
4.44444444444444 1.04651162790698
4.34343434343434 1.0454065469905
4.24242424242424 1.04430379746835
4.14141414141414 1.0432033719705
4.04040404040404 1.04210526315789
3.93939393939394 1.0410094637224
3.83838383838384 1.03991596638655
3.73737373737374 1.03882476390346
3.63636363636364 1.0377358490566
3.53535353535354 1.03664921465969
3.43434343434343 1.03556485355649
3.33333333333333 1.03448275862069
3.23232323232323 1.03340292275574
3.13131313131313 1.03232533889468
3.03030303030303 1.03125
2.92929292929293 1.03017689906348
2.82828282828283 1.02910602910603
2.72727272727273 1.02803738317757
2.62626262626263 1.02697095435685
2.52525252525253 1.0259067357513
2.42424242424242 1.02484472049689
2.32323232323232 1.02378490175801
2.22222222222222 1.02272727272727
2.12121212121212 1.02167182662539
2.02020202020202 1.02061855670103
1.91919191919192 1.01956745623069
1.81818181818182 1.01851851851852
1.71717171717172 1.0174717368962
1.61616161616162 1.01642710472279
1.51515151515152 1.01538461538462
1.41414141414141 1.01434426229508
1.31313131313131 1.01330603889458
1.21212121212121 1.01226993865031
1.11111111111111 1.01123595505618
1.01010101010101 1.01020408163265
0.90909090909091 1.00917431192661
0.808080808080808 1.0081466395112
0.707070707070708 1.00712105798576
0.606060606060606 1.00609756097561
0.505050505050505 1.00507614213198
0.404040404040405 1.00405679513185
0.303030303030303 1.00303951367781
0.202020202020202 1.00202429149798
0.1010101010101 1.0010111223458
0 1
-0.222222222222222 0.997782705099778
-0.444444444444444 0.995575221238938
-0.666666666666667 0.993377483443709
-0.888888888888889 0.991189427312775
-1.11111111111111 0.989010989010989
-1.33333333333333 0.986842105263158
-1.55555555555556 0.984682713347921
-1.77777777777778 0.982532751091703
-2 0.980392156862745
};
\addplot [ultra thick, blue]
table {%
10 1.05557944270438
9.8989898989899 1.03446856464943
9.7979797979798 1.01377988994008
9.6969696969697 0.993504974793709
9.5959595959596 0.973635544297678
9.49494949494949 0.954163489032021
9.39393939393939 0.935080861759713
9.29292929292929 0.916379874183118
9.19191919191919 0.898052893765306
9.09090909090909 0.880092440614949
8.98989898989899 0.862491184433507
8.88888888888889 0.845241941523476
8.78787878787879 0.828337671856464
8.68686868686869 0.811771476199906
8.58585858585859 0.795536593301243
8.48484848484848 0.779626397128416
8.38383838383838 0.764034394165541
8.28282828282828 0.748754220762684
8.18181818181818 0.733779640538621
8.08080808080808 0.719104541835555
7.97979797979798 0.704722935224729
7.87878787878788 0.690628951061928
7.77777777777778 0.676816837091868
7.67676767676768 0.6632809561005
7.57575757575758 0.650015783614256
7.47474747474747 0.637015905645323
7.37373737373737 0.624276016481996
7.27272727272727 0.611790916523232
7.17171717171717 0.59955551015651
7.07070707070707 0.587564803678124
6.96969696969697 0.575813903255083
6.86868686868687 0.564298012927759
6.76767676767677 0.553012432652488
6.66666666666667 0.541952556383311
6.56565656565657 0.531113870192092
6.46464646464646 0.520491950426213
6.36363636363636 0.510082461903134
6.26262626262626 0.499881156141044
6.16161616161616 0.489883869624907
6.06060606060606 0.480086522107185
5.95959595959596 0.470485114942545
5.85858585858586 0.461075729455866
5.75757575757576 0.451854525342889
5.65656565656566 0.442817739102859
5.55555555555556 0.433961682502498
5.45454545454546 0.42528274107071
5.35353535353535 0.41677737262339
5.25252525252525 0.408442105817727
5.15151515151515 0.400273538735431
5.05050505050505 0.392268337494284
4.94949494949495 0.38442323488747
4.84848484848485 0.37673502905011
4.74747474747475 0.369200582152463
4.64646464646465 0.361816819119272
4.54545454545455 0.354580726374716
4.44444444444444 0.347489350612458
4.34343434343434 0.340539797590302
4.24242424242424 0.333729230948944
4.14141414141414 0.327054871054359
4.04040404040404 0.320513993863329
3.93939393939394 0.314103929811667
3.83838383838384 0.307822062724671
3.73737373737374 0.301665828749373
3.63636363636364 0.295632715308137
3.53535353535354 0.289720260073187
3.43434343434343 0.283926049961645
3.33333333333333 0.278247720150668
3.23232323232323 0.272682953112274
3.13131313131313 0.267229477667481
3.03030303030303 0.261885068059359
2.92929292929293 0.256647543044616
2.82828282828283 0.251514765003358
2.72727272727273 0.246484639066648
2.62626262626263 0.24155511226152
2.52525252525253 0.236724172673081
2.42424242424242 0.231989848623386
2.32323232323232 0.227350207866716
2.22222222222222 0.222803356800969
2.12121212121212 0.218347439694808
2.02020202020202 0.213980637930274
1.91919191919192 0.209701169260543
1.81818181818182 0.205507287082525
1.71717171717172 0.20139727972402
1.61616161616162 0.197369469745115
1.51515151515152 0.193422213253567
1.41414141414141 0.189553899233873
1.31313131313131 0.185762948889753
1.21212121212121 0.182047814999791
1.11111111111111 0.178406981285955
1.01010101010101 0.174838961794755
0.90909090909091 0.171342300290769
0.808080808080808 0.167915569662305
0.707070707070708 0.16455737133894
0.606060606060606 0.161266334720721
0.505050505050505 0.158041116618771
0.404040404040405 0.154880400707082
0.303030303030303 0.151782896985286
0.202020202020202 0.148747341252149
0.1010101010101 0.145772494589612
0 0.142857142857143
-0.222222222222222 0.136646962729004
-0.444444444444444 0.130706746961433
-0.666666666666667 0.125024759863278
-0.888888888888889 0.119589775908681
-1.11111111111111 0.114391057559544
-1.33333333333333 0.109418334052093
-1.55555555555556 0.104661781105602
-1.77777777777778 0.100112001513219
-2 0.0957600065765199
};
\addplot [thick, red, dashed]
table {%
10 1.09877913429523
9.8989898989899 1.09877913429523
9.7979797979798 1.09877913429523
9.6969696969697 1.09877913429523
9.5959595959596 1.09877913429523
9.49494949494949 1.09877913429523
9.39393939393939 1.09877913429523
9.29292929292929 1.09877913429523
9.19191919191919 1.09877913429523
9.09090909090909 1.09877913429523
8.98989898989899 1.09877913429523
8.88888888888889 1.09877913429523
8.78787878787879 1.09877913429523
8.68686868686869 1.09877913429523
8.58585858585859 1.09877913429523
8.48484848484848 1.09877913429523
8.38383838383838 1.09877913429523
8.28282828282828 1.09877913429523
8.18181818181818 1.09877913429523
8.08080808080808 1.09877913429523
7.97979797979798 1.09877913429523
7.87878787878788 1.09877913429523
7.77777777777778 1.09877913429523
7.67676767676768 1.09877913429523
7.57575757575758 1.09877913429523
7.47474747474747 1.09877913429523
7.37373737373737 1.09877913429523
7.27272727272727 1.09877913429523
7.17171717171717 1.09877913429523
7.07070707070707 1.09877913429523
6.96969696969697 1.09877913429523
6.86868686868687 1.09877913429523
6.76767676767677 1.09877913429523
6.66666666666667 1.09877913429523
6.56565656565657 1.09877913429523
6.46464646464646 1.09877913429523
6.36363636363636 1.09877913429523
6.26262626262626 1.09877913429523
6.16161616161616 1.09877913429523
6.06060606060606 1.09877913429523
5.95959595959596 1.09877913429523
5.85858585858586 1.09877913429523
5.75757575757576 1.09877913429523
5.65656565656566 1.09877913429523
5.55555555555556 1.09877913429523
5.45454545454546 1.09877913429523
5.35353535353535 1.09877913429523
5.25252525252525 1.09877913429523
5.15151515151515 1.09877913429523
5.05050505050505 1.09877913429523
4.94949494949495 1.09877913429523
4.84848484848485 1.09877913429523
4.74747474747475 1.09877913429523
4.64646464646465 1.09877913429523
4.54545454545455 1.09877913429523
4.44444444444444 1.09877913429523
4.34343434343434 1.09877913429523
4.24242424242424 1.09877913429523
4.14141414141414 1.09877913429523
4.04040404040404 1.09877913429523
3.93939393939394 1.09877913429523
3.83838383838384 1.09877913429523
3.73737373737374 1.09877913429523
3.63636363636364 1.09877913429523
3.53535353535354 1.09877913429523
3.43434343434343 1.09877913429523
3.33333333333333 1.09877913429523
3.23232323232323 1.09877913429523
3.13131313131313 1.09877913429523
3.03030303030303 1.09877913429523
2.92929292929293 1.09877913429523
2.82828282828283 1.09877913429523
2.72727272727273 1.09877913429523
2.62626262626263 1.09877913429523
2.52525252525253 1.09877913429523
2.42424242424242 1.09877913429523
2.32323232323232 1.09877913429523
2.22222222222222 1.09877913429523
2.12121212121212 1.09877913429523
2.02020202020202 1.09877913429523
1.91919191919192 1.09877913429523
1.81818181818182 1.09877913429523
1.71717171717172 1.09877913429523
1.61616161616162 1.09877913429523
1.51515151515152 1.09877913429523
1.41414141414141 1.09877913429523
1.31313131313131 1.09877913429523
1.21212121212121 1.09877913429523
1.11111111111111 1.09877913429523
1.01010101010101 1.09877913429523
0.90909090909091 1.09877913429523
0.808080808080808 1.09877913429523
0.707070707070708 1.09877913429523
0.606060606060606 1.09877913429523
0.505050505050505 1.09877913429523
0.404040404040405 1.09877913429523
0.303030303030303 1.09877913429523
0.202020202020202 1.09877913429523
0.1010101010101 1.09877913429523
0 1.09877913429523
-0.222222222222222 1.09877913429523
-0.444444444444444 1.09877913429523
-0.666666666666667 1.09877913429523
-0.888888888888889 1.09877913429523
-1.11111111111111 1.09877913429523
-1.33333333333333 1.09877913429523
-1.55555555555556 1.09877913429523
-1.77777777777778 1.09877913429523
-2 1.09877913429523
};
\addplot [thick, red, dashed]
table {%
10 0.862491184433507
9.8989898989899 0.862491184433507
9.7979797979798 0.862491184433507
9.6969696969697 0.862491184433507
9.5959595959596 0.862491184433507
9.49494949494949 0.862491184433507
9.39393939393939 0.862491184433507
9.29292929292929 0.862491184433507
9.19191919191919 0.862491184433507
9.09090909090909 0.862491184433507
8.98989898989899 0.862491184433507
8.88888888888889 0.862491184433507
8.78787878787879 0.862491184433507
8.68686868686869 0.862491184433507
8.58585858585859 0.862491184433507
8.48484848484848 0.862491184433507
8.38383838383838 0.862491184433507
8.28282828282828 0.862491184433507
8.18181818181818 0.862491184433507
8.08080808080808 0.862491184433507
7.97979797979798 0.862491184433507
7.87878787878788 0.862491184433507
7.77777777777778 0.862491184433507
7.67676767676768 0.862491184433507
7.57575757575758 0.862491184433507
7.47474747474747 0.862491184433507
7.37373737373737 0.862491184433507
7.27272727272727 0.862491184433507
7.17171717171717 0.862491184433507
7.07070707070707 0.862491184433507
6.96969696969697 0.862491184433507
6.86868686868687 0.862491184433507
6.76767676767677 0.862491184433507
6.66666666666667 0.862491184433507
6.56565656565657 0.862491184433507
6.46464646464646 0.862491184433507
6.36363636363636 0.862491184433507
6.26262626262626 0.862491184433507
6.16161616161616 0.862491184433507
6.06060606060606 0.862491184433507
5.95959595959596 0.862491184433507
5.85858585858586 0.862491184433507
5.75757575757576 0.862491184433507
5.65656565656566 0.862491184433507
5.55555555555556 0.862491184433507
5.45454545454546 0.862491184433507
5.35353535353535 0.862491184433507
5.25252525252525 0.862491184433507
5.15151515151515 0.862491184433507
5.05050505050505 0.862491184433507
4.94949494949495 0.862491184433507
4.84848484848485 0.862491184433507
4.74747474747475 0.862491184433507
4.64646464646465 0.862491184433507
4.54545454545455 0.862491184433507
4.44444444444444 0.862491184433507
4.34343434343434 0.862491184433507
4.24242424242424 0.862491184433507
4.14141414141414 0.862491184433507
4.04040404040404 0.862491184433507
3.93939393939394 0.862491184433507
3.83838383838384 0.862491184433507
3.73737373737374 0.862491184433507
3.63636363636364 0.862491184433507
3.53535353535354 0.862491184433507
3.43434343434343 0.862491184433507
3.33333333333333 0.862491184433507
3.23232323232323 0.862491184433507
3.13131313131313 0.862491184433507
3.03030303030303 0.862491184433507
2.92929292929293 0.862491184433507
2.82828282828283 0.862491184433507
2.72727272727273 0.862491184433507
2.62626262626263 0.862491184433507
2.52525252525253 0.862491184433507
2.42424242424242 0.862491184433507
2.32323232323232 0.862491184433507
2.22222222222222 0.862491184433507
2.12121212121212 0.862491184433507
2.02020202020202 0.862491184433507
1.91919191919192 0.862491184433507
1.81818181818182 0.862491184433507
1.71717171717172 0.862491184433507
1.61616161616162 0.862491184433507
1.51515151515152 0.862491184433507
1.41414141414141 0.862491184433507
1.31313131313131 0.862491184433507
1.21212121212121 0.862491184433507
1.11111111111111 0.862491184433507
1.01010101010101 0.862491184433507
0.90909090909091 0.862491184433507
0.808080808080808 0.862491184433507
0.707070707070708 0.862491184433507
0.606060606060606 0.862491184433507
0.505050505050505 0.862491184433507
0.404040404040405 0.862491184433507
0.303030303030303 0.862491184433507
0.202020202020202 0.862491184433507
0.1010101010101 0.862491184433507
0 0.862491184433507
-0.222222222222222 0.862491184433507
-0.444444444444444 0.862491184433507
-0.666666666666667 0.862491184433507
-0.888888888888889 0.862491184433507
-1.11111111111111 0.862491184433507
-1.33333333333333 0.862491184433507
-1.55555555555556 0.862491184433507
-1.77777777777778 0.862491184433507
-2 0.862491184433507
};
\addplot [thick, red, dashed]
table {%
10 1.01123595505618
9.8989898989899 1.01123595505618
9.7979797979798 1.01123595505618
9.6969696969697 1.01123595505618
9.5959595959596 1.01123595505618
9.49494949494949 1.01123595505618
9.39393939393939 1.01123595505618
9.29292929292929 1.01123595505618
9.19191919191919 1.01123595505618
9.09090909090909 1.01123595505618
8.98989898989899 1.01123595505618
8.88888888888889 1.01123595505618
8.78787878787879 1.01123595505618
8.68686868686869 1.01123595505618
8.58585858585859 1.01123595505618
8.48484848484848 1.01123595505618
8.38383838383838 1.01123595505618
8.28282828282828 1.01123595505618
8.18181818181818 1.01123595505618
8.08080808080808 1.01123595505618
7.97979797979798 1.01123595505618
7.87878787878788 1.01123595505618
7.77777777777778 1.01123595505618
7.67676767676768 1.01123595505618
7.57575757575758 1.01123595505618
7.47474747474747 1.01123595505618
7.37373737373737 1.01123595505618
7.27272727272727 1.01123595505618
7.17171717171717 1.01123595505618
7.07070707070707 1.01123595505618
6.96969696969697 1.01123595505618
6.86868686868687 1.01123595505618
6.76767676767677 1.01123595505618
6.66666666666667 1.01123595505618
6.56565656565657 1.01123595505618
6.46464646464646 1.01123595505618
6.36363636363636 1.01123595505618
6.26262626262626 1.01123595505618
6.16161616161616 1.01123595505618
6.06060606060606 1.01123595505618
5.95959595959596 1.01123595505618
5.85858585858586 1.01123595505618
5.75757575757576 1.01123595505618
5.65656565656566 1.01123595505618
5.55555555555556 1.01123595505618
5.45454545454546 1.01123595505618
5.35353535353535 1.01123595505618
5.25252525252525 1.01123595505618
5.15151515151515 1.01123595505618
5.05050505050505 1.01123595505618
4.94949494949495 1.01123595505618
4.84848484848485 1.01123595505618
4.74747474747475 1.01123595505618
4.64646464646465 1.01123595505618
4.54545454545455 1.01123595505618
4.44444444444444 1.01123595505618
4.34343434343434 1.01123595505618
4.24242424242424 1.01123595505618
4.14141414141414 1.01123595505618
4.04040404040404 1.01123595505618
3.93939393939394 1.01123595505618
3.83838383838384 1.01123595505618
3.73737373737374 1.01123595505618
3.63636363636364 1.01123595505618
3.53535353535354 1.01123595505618
3.43434343434343 1.01123595505618
3.33333333333333 1.01123595505618
3.23232323232323 1.01123595505618
3.13131313131313 1.01123595505618
3.03030303030303 1.01123595505618
2.92929292929293 1.01123595505618
2.82828282828283 1.01123595505618
2.72727272727273 1.01123595505618
2.62626262626263 1.01123595505618
2.52525252525253 1.01123595505618
2.42424242424242 1.01123595505618
2.32323232323232 1.01123595505618
2.22222222222222 1.01123595505618
2.12121212121212 1.01123595505618
2.02020202020202 1.01123595505618
1.91919191919192 1.01123595505618
1.81818181818182 1.01123595505618
1.71717171717172 1.01123595505618
1.61616161616162 1.01123595505618
1.51515151515152 1.01123595505618
1.41414141414141 1.01123595505618
1.31313131313131 1.01123595505618
1.21212121212121 1.01123595505618
1.11111111111111 1.01123595505618
1.01010101010101 1.01123595505618
0.90909090909091 1.01123595505618
0.808080808080808 1.01123595505618
0.707070707070708 1.01123595505618
0.606060606060606 1.01123595505618
0.505050505050505 1.01123595505618
0.404040404040405 1.01123595505618
0.303030303030303 1.01123595505618
0.202020202020202 1.01123595505618
0.1010101010101 1.01123595505618
0 1.01123595505618
-0.222222222222222 1.01123595505618
-0.444444444444444 1.01123595505618
-0.666666666666667 1.01123595505618
-0.888888888888889 1.01123595505618
-1.11111111111111 1.01123595505618
-1.33333333333333 1.01123595505618
-1.55555555555556 1.01123595505618
-1.77777777777778 1.01123595505618
-2 1.01123595505618
};
\addplot [thick, red, dashed]
table {%
10 0.178406981285955
9.8989898989899 0.178406981285955
9.7979797979798 0.178406981285955
9.6969696969697 0.178406981285955
9.5959595959596 0.178406981285955
9.49494949494949 0.178406981285955
9.39393939393939 0.178406981285955
9.29292929292929 0.178406981285955
9.19191919191919 0.178406981285955
9.09090909090909 0.178406981285955
8.98989898989899 0.178406981285955
8.88888888888889 0.178406981285955
8.78787878787879 0.178406981285955
8.68686868686869 0.178406981285955
8.58585858585859 0.178406981285955
8.48484848484848 0.178406981285955
8.38383838383838 0.178406981285955
8.28282828282828 0.178406981285955
8.18181818181818 0.178406981285955
8.08080808080808 0.178406981285955
7.97979797979798 0.178406981285955
7.87878787878788 0.178406981285955
7.77777777777778 0.178406981285955
7.67676767676768 0.178406981285955
7.57575757575758 0.178406981285955
7.47474747474747 0.178406981285955
7.37373737373737 0.178406981285955
7.27272727272727 0.178406981285955
7.17171717171717 0.178406981285955
7.07070707070707 0.178406981285955
6.96969696969697 0.178406981285955
6.86868686868687 0.178406981285955
6.76767676767677 0.178406981285955
6.66666666666667 0.178406981285955
6.56565656565657 0.178406981285955
6.46464646464646 0.178406981285955
6.36363636363636 0.178406981285955
6.26262626262626 0.178406981285955
6.16161616161616 0.178406981285955
6.06060606060606 0.178406981285955
5.95959595959596 0.178406981285955
5.85858585858586 0.178406981285955
5.75757575757576 0.178406981285955
5.65656565656566 0.178406981285955
5.55555555555556 0.178406981285955
5.45454545454546 0.178406981285955
5.35353535353535 0.178406981285955
5.25252525252525 0.178406981285955
5.15151515151515 0.178406981285955
5.05050505050505 0.178406981285955
4.94949494949495 0.178406981285955
4.84848484848485 0.178406981285955
4.74747474747475 0.178406981285955
4.64646464646465 0.178406981285955
4.54545454545455 0.178406981285955
4.44444444444444 0.178406981285955
4.34343434343434 0.178406981285955
4.24242424242424 0.178406981285955
4.14141414141414 0.178406981285955
4.04040404040404 0.178406981285955
3.93939393939394 0.178406981285955
3.83838383838384 0.178406981285955
3.73737373737374 0.178406981285955
3.63636363636364 0.178406981285955
3.53535353535354 0.178406981285955
3.43434343434343 0.178406981285955
3.33333333333333 0.178406981285955
3.23232323232323 0.178406981285955
3.13131313131313 0.178406981285955
3.03030303030303 0.178406981285955
2.92929292929293 0.178406981285955
2.82828282828283 0.178406981285955
2.72727272727273 0.178406981285955
2.62626262626263 0.178406981285955
2.52525252525253 0.178406981285955
2.42424242424242 0.178406981285955
2.32323232323232 0.178406981285955
2.22222222222222 0.178406981285955
2.12121212121212 0.178406981285955
2.02020202020202 0.178406981285955
1.91919191919192 0.178406981285955
1.81818181818182 0.178406981285955
1.71717171717172 0.178406981285955
1.61616161616162 0.178406981285955
1.51515151515152 0.178406981285955
1.41414141414141 0.178406981285955
1.31313131313131 0.178406981285955
1.21212121212121 0.178406981285955
1.11111111111111 0.178406981285955
1.01010101010101 0.178406981285955
0.90909090909091 0.178406981285955
0.808080808080808 0.178406981285955
0.707070707070708 0.178406981285955
0.606060606060606 0.178406981285955
0.505050505050505 0.178406981285955
0.404040404040405 0.178406981285955
0.303030303030303 0.178406981285955
0.202020202020202 0.178406981285955
0.1010101010101 0.178406981285955
0 0.178406981285955
-0.222222222222222 0.178406981285955
-0.444444444444444 0.178406981285955
-0.666666666666667 0.178406981285955
-0.888888888888889 0.178406981285955
-1.11111111111111 0.178406981285955
-1.33333333333333 0.178406981285955
-1.55555555555556 0.178406981285955
-1.77777777777778 0.178406981285955
-2 0.178406981285955
};
\addplot [thick, red, densely dashed,name path = C]
table { 
1.11111111111111 1e-03
1.11111111111111 5e+01
};
\addplot [thick, red, densely dashed,name path = D]
table { 
8.98989898989899 1e-03
8.98989898989899 5e+01
};
\addplot [thick, black, <->]
table { 
6 1.09877913429523
6 1.01123595505618
};
\addplot [thick, black, <->]
table { 
6 0.862491184433507
6 0.178406981285955
};
\addplot [thick, red, densely dashed,name path = E]
table { 
-10.1 1e-03
-10.1 5e+01
};
\addplot[mark=*, red,mark size=3, mark options={solid}] coordinates {(1.11111111111111,0.178406981285955)};
\addplot[mark=*, red,mark size=3, mark options={solid}] coordinates {(1.11111111111111,1.01123595505618)};
\addplot[mark=*, red,mark size=3, mark options={solid}] coordinates {(8.98989898989899,0.862491184433507)};
\addplot[mark=*, red,mark size=3, mark options={solid}] coordinates {(8.98989898989899,1.09877913429523)};
\addplot[domain=0:8.98989898989899,name path=B,update limits=false] {-0.05};
\addplot[opacity=0.5,draw=none, fill=red!25] fill between[of=C and D];
\addplot[opacity=0.5,draw=none, fill=black!25] fill between[of=D and E];

\end{axis}

\begin{axis}[
axis y line*=right,
  axis x line=none,
x dir=reverse,
xmin=-2, xmax=10,
ymin=0.0449924513497903, ymax=1.16187866633784,
xtick={8.98989898989899,1.11111111111111},
xticklabels={$\lambda_1$,$\lambda_2$},
ytick={0.862491184433507,0.178406981285955},
yticklabels={$h_{\sf s}(\lambda_1)$,\small $h_{\sf s}(\lambda_2)$},
width=6cm,height=5cm,
]
\end{axis}

\end{tikzpicture}

%% file: legend_tikz.tex
\begin{tikzpicture}
    \begin{customlegend}[legend columns=4, legend style={column sep=0.1ex, font=\normalsize, cells={align=left}},legend entries={RPCA \eqref{eq:rpca},Algorithm~\ref{alg:mega},GL-SigRep \cite{dong2016learning}, KNN, PCA \eqref{eq:pca}, SpecTemp \cite{segarra2017network}, Kalofolias \cite{kalofolias2016learn}}]
    \addlegendimage{very thick, dashed, blue, mark=o, mark size=4, mark options={solid}, sharp plot}
    \addlegendimage{very thick, red, mark=square, mark size=4, mark options={solid}, sharp plot}
    \addlegendimage{thick, color0, mark=diamond, mark size=3, mark options={solid}, sharp plot}
    \addlegendimage{thick, green!50!black, mark=x, mark size=4, mark options={solid}, sharp plot}
    \addlegendimage{very thick, black, mark=otimes, mark size=4, mark options={solid},sharp plot}
    \addlegendimage{thick, orange, mark=triangle, mark size=4, mark options={solid}, sharp plot}
    \addlegendimage{thick, magenta, mark=oplus, mark size=4, mark options={solid}, sharp plot}
    \end{customlegend}
    \end{tikzpicture}

%% file: Evsk_wlp_cp.tex
\begin{tikzpicture}

\definecolor{color0}{rgb}{0,0.75,0.75}

\begin{axis}[
legend cell align={left},
legend columns=3,
legend style={at={(0.5,1.05)}, anchor=north, draw=white!80.0!black},
tick align=outside,
tick pos=both,
title={},
x grid style={white!69.01960784313725!black},
xlabel={\Large Excitation's Rank $k$},
xmajorgrids,
xmin=8, xmax=52,
xtick style={color=black},
y grid style={white!69.01960784313725!black},
ylabel={\Large Error rate},
ymajorgrids,
ymin=0, ymax=1.05,
ytick style={color=black},
width=7cm,height=8cm,
]
\addplot [very thick, dashed, blue, mark=o, mark size=4, mark options={solid}]
table {%
10 0.78
20 0.475
30 0.219
40 0.132
50 0.0979999999999999
};
\addplot [very thick, red, mark=square, mark size=4, mark options={solid}]
table {%
10 0.782
20 0.438
30 0.24
40 0.191
50 0.308
};
\addplot [thick, color0, mark=diamond, mark size=3, mark options={solid}]
table {%
10 0.759
20 0.847
30 0.896
40 0.951
50 0.976
};
\addplot [thick, green!50!black, mark=x, mark size=4, mark options={solid}]
table {%
10 0.528
20 0.71
30 0.754
40 0.725
50 0.726
};
\addplot [very thick, black, mark=otimes, mark size=4, mark options={solid}]
table {%
10 0.792
20 0.744
30 0.694
40 0.652
50 0.588
};

\addplot [thick, orange, mark=triangle, mark size=4, mark options={solid}]
table {%
10 0.761
20 0.71
30 0.638
40 0.59
50 0.528
};

\addplot [thick, magenta, mark=oplus, mark size=4, mark options={solid}]
table {%
10 1.0000
20 0.9920
30 1.0000
40 0.9940
50 0.9800
};
\end{axis}

\end{tikzpicture}

%% file: Evsk_slp_cp.tex

\begin{tikzpicture}    [
     spy using outlines={circle,magnification=2.65, size=0.5cm, connect spies}]

\definecolor{color0}{rgb}{0,0.75,0.75}

\begin{axis}[
legend cell align={left},
legend columns=1,
legend style={at={(0.5,0.7)}, anchor=north, draw=white!80.0!black},
tick align=outside,
tick pos=both,
title={},
x grid style={white!69.01960784313725!black},
xlabel={\Large Excitation's Rank $k$},
xmajorgrids,
xmin=8, xmax=52,
xtick style={color=black},
y grid style={white!69.01960784313725!black},
ymajorgrids,
ymin=0, ymax=1.05,
ytick style={color=black},
width=7cm,height=8cm,
]
\addplot [thick, dashed, blue, mark=o, mark size=4, mark options={solid}]
table {%
10 0.07
20 0.006
30 0.005
40 0.002
50 0.002
};
\addplot [thick, red, mark=square, mark size=4, mark options={solid}]
table {%
10 0.063
20 0.008
30 0.006
40 0.002
50 0.002
};
\addplot [thick, color0, mark=diamond, mark size=3, mark options={solid}]
table {%
10 0.99
20 1
30 1
40 1
50 1
};
\addplot [thick, green!50!black, mark=x, mark size=4, mark options={solid}]
table {%
10 0.999
20 1
30 1
40 1
50 1
};
\addplot [thick, black, mark=otimes, mark size=4, mark options={solid}]
table {%
10 0.084
20 0.023
30 0.019
40 0.009
50 0.011
};

\addplot [thick, orange, mark=triangle, mark size=4, mark options={solid}]
table {%
10 0.071
20 0.025
30 0.007
40 0.006
50 0.008
};

\addplot [thick, magenta, mark=oplus, mark size=4, mark options={solid}]
table {%
10 1.0000
20 1.0000
30 0.9960
40 0.9940
50 0.9380
};
\begin{scope}
    \spy[blue,width=4cm,height=2cm] on (0.8,0.5) in node [fill=white] at (3.0,3.0);
\end{scope}
\end{axis}

\end{tikzpicture}

%% file: Evsp_wlp_cp.tex
\begin{tikzpicture}

\definecolor{color0}{rgb}{0,0.75,0.75}

\begin{axis}[
legend cell align={left},
legend columns=1,
legend style={at={(0.5,1)}, anchor=north, draw=white!80.0!black},
tick align=outside,
tick pos=both,
title={},
x grid style={white!69.01960784313725!black},
xlabel={\large $p_1$},
xmajorgrids,
xmin=0.1, xmax=1,
xtick style={color=black},
y grid style={white!69.01960784313725!black},
ylabel={\large Error rate},
ymajorgrids,
ymin=0, ymax=1.05,
ytick style={color=black},
width=7cm,height=8cm,
]
\addplot [very thick, dashed, blue, mark=o, mark size=4, mark options={solid}]
table {%
0.1 0.263
0.2 0.23
0.4 0.132
0.6 0.092
0.8 0.061
1 0.05
};
\addplot [very thick, red, mark=square, mark size=4, mark options={solid}]
table {%
0.1 0.322
0.2 0.277
0.4 0.186
0.6 0.145
0.8 0.1355
1 0.125
};
\addplot [thick, color0, mark=diamond, mark size=3, mark options={solid}]
table {%
0.1 0.952
0.2 0.932
0.4 0.957
0.6 0.974
0.8 0.965
1 0.97
};
\addplot [thick, green!50!black, mark=x, mark size=4, mark options={solid}]
table {%
0.1 0.752
0.2 0.712
0.4 0.777
0.6 0.751
0.8 0.774
1 0.75
};
\addplot [thick, black, mark=otimes, mark size=4, mark options={solid}]
table {%
0.1 0.707
0.2 0.7
0.4 0.639
0.6 0.609
0.8 0.576
1 0.587
};

\addplot [thick, orange, mark=triangle, mark size=4, mark options={solid}]
table {%
0.1 0.6900
0.2 0.6740
0.4 0.6120
0.6 0.5520
0.8 0.5260
1 0.5040
};

\addplot [thick, magenta, mark=oplus, mark size=4, mark options={solid}]
table {%
0.1 0.9840
0.2 0.9900
0.4 0.9980
0.6 0.9980
0.8 0.9920
1 1.0000
};

\end{axis}

\end{tikzpicture}

%% file: Evsp_slp_cp.tex
\begin{tikzpicture}[
     spy using outlines={circle,magnification=2.65, size=0.5cm, connect spies}]

\definecolor{color0}{rgb}{0,0.75,0.75}

\begin{axis}[
legend cell align={left},
legend columns=1,
legend style={at={(0.5,1)}, anchor=north, draw=white!80.0!black},
tick align=outside,
tick pos=both,
title={},
x grid style={white!69.01960784313725!black},
xlabel={\large $p_1$},
xmajorgrids,
xmin=0.1, xmax=1,
xtick style={color=black},
y grid style={white!69.01960784313725!black},
ymajorgrids,
ymin=0, ymax=1.05,
ytick style={color=black},
width=7cm,height=8cm,
]
\addplot [very thick, dashed, blue, mark=o, mark size=4, mark options={solid}]
table {%
0.1 0.032
0.2 0.018
0.4 0.004
0.6 0
0.8 0
1 0
};
\addplot [very thick, red, mark=square, mark size=4, mark options={solid}]
table {%
0.1 0.042
0.2 0.034
0.4 0.006
0.6 0.002
0.8 0
1 0
};
\addplot [thick, color0, mark=diamond, mark size=3, mark options={solid}]
table {%
0.1 1
0.2 1
0.4 1
0.6 1
0.8 1
1 1
};
\addplot [thick, green!50!black, mark=x, mark size=4, mark options={solid}]
table {%
0.1 1
0.2 1
0.4 1
0.6 1
0.8 1
1 1
};
\addplot [thick, black, mark=otimes, mark size=4, mark options={solid}]
table {%
0.1 0.072
0.2 0.05
0.4 0.008
0.6 0
0.8 0
1 0
};

\addplot [thick, orange, mark=triangle, mark size=4, mark options={solid}]
table {%
0.1 0.0960
0.2 0.0460
0.4 0.0040
0.6 0.0060
0.8 0
1 0.0020
};

\addplot [thick, magenta, mark=oplus, mark size=4, mark options={solid}]
table {%
0.1 0.9900
0.2 1.0000
0.4 0.9960
0.6 1.0000
0.8 0.9920
1 1.0000
};
\begin{scope}
    \spy[blue,width=4cm,height=2cm] on (0.5,0.5) in node [fill=white] at (3.0,3.0);
\end{scope}

\end{axis}

\end{tikzpicture}

%% file: Evsk_wkp_ba.tex
\begin{tikzpicture}

\definecolor{color0}{rgb}{0,0.75,0.75}

\begin{axis}[
legend cell align={left},
legend columns=2,
legend style={at={(0.5,0.7)}, anchor=north, draw=white!80.0!black},
tick align=outside,
tick pos=both,
title={},
x grid style={white!69.01960784313725!black},
xlabel={\Large Excitation's Rank $k$},
xmajorgrids,
xmin=8, xmax=52,
xtick style={color=black},
y grid style={white!69.01960784313725!black},
ylabel={\Large Error rate},
ymajorgrids,
ymin=0.15, ymax=0.7,
ytick style={color=black},
width=7cm,height=8cm,
]
\addplot [very thick, dashed, blue, mark=o, mark size=4, mark options={solid}]
table {%
10 0.3976
20 0.2568
30 0.2162
40 0.19
50 0.1772
};
\addplot [very thick, red, mark=square, mark size=4, mark options={solid}]
table {%
10 0.4082
20 0.2924
30 0.2724
40 0.2686
50 0.2712
};
\addplot [thick, color0, mark=diamond, mark size=3, mark options={solid}]
table {%
10 0.4302
20 0.4884
30 0.5426
40 0.5832
50 0.6286
};
\addplot [thick, green!50!black, mark=x, mark size=4, mark options={solid}]
table {%
10 0.3936
20 0.4484
30 0.459
40 0.4494
50 0.4574
};
\addplot [thick, black, mark=otimes, mark size=4, mark options={solid}]
table {%
10 0.4038
20 0.386
30 0.382
40 0.3582
50 0.3456
};

\addplot [thick, orange, mark=triangle, mark size=4, mark options={solid}]
table {%
10 0.4408
20 0.3726
30 0.372
40 0.3484
50 0.3128
};

\addplot [thick, magenta, mark=oplus, mark size=4, mark options={solid}]
table {%
10 0.5152
20 0.5060
30 0.4824
40 0.4852
50 0.4984
};

\end{axis}

\end{tikzpicture}

%% file: Evsk_slp_ba.tex
\begin{tikzpicture} [
     spy using outlines={circle,magnification=2.65, size=0.5cm, connect spies}]

\definecolor{color0}{rgb}{0,0.75,0.75}

\begin{axis}[
legend cell align={left},
legend columns=2,
legend style={at={(0.5,0.5)}, anchor=north, draw=white!80.0!black},
tick align=outside,
tick pos=both,
title={},
x grid style={white!69.01960784313725!black},
xlabel={\Large Excitation's Rank $k$},
xmajorgrids,
xmin=8, xmax=52,
xtick style={color=black},
y grid style={white!69.01960784313725!black},
ymajorgrids,
ymin=0, ymax=1,
ytick style={color=black},
width=7cm,height=8cm,
]
\addplot [very thick, dashed, blue, mark=o, mark size=4, mark options={solid}]
table {%
10 0.1486
20 0.0641
30 0.0497
40 0.047
50 0.0465
};
\addplot [very thick, red, mark=square, mark size=4, mark options={solid}]
table {%
10 0.1544
20 0.0973
30 0.0835
40 0.0777
50 0.0674
};
\addplot [thick, color0, mark=diamond, mark size=3, mark options={solid}]
table {%
10 0.5698
20 0.7814
30 0.8427
40 0.8598
50 0.8439
};
\addplot [thick, green!50!black, mark=x, mark size=4, mark options={solid}]
table {%
10 0.58
20 0.7597
30 0.8197
40 0.8183
50 0.8574
};
\addplot [very thick, black, mark=otimes, mark size=4, mark options={solid}]
table {%
10 0.1543
20 0.1155
30 0.0949
40 0.0891
50 0.0781
};

\addplot [thick, orange, mark=triangle, mark size=4, mark options={solid}]
table {%
10 0.0932
20 0.0756
30 0.0642
40 0.0646
50 0.0582
};

\addplot [thick, magenta, mark=oplus, mark size=4, mark options={solid}]
table {%
10 0.5784
20 0.5460
30 0.5328
40 0.5300
50 0.4888
};
\begin{scope}
    \spy[blue,width=2cm,height=2.5cm] on (1.5,0.5) in node [fill=white] at (3.0,2.0);
\end{scope}
\end{axis}

\end{tikzpicture}

%% file: Evsn_wlp_cp.tex
\begin{tikzpicture}

\definecolor{color0}{rgb}{0,0.75,0.75}

\begin{axis}[
legend cell align={left},
legend columns=1,
legend style={at={(0.5,1)}, anchor=north, draw=white!80.0!black},
tick align=outside,
tick pos=both,
title={},
x grid style={white!69.01960784313725!black},
xlabel={\Large Graph size $n$},
xmajorgrids,
xmin=80, xmax=160,
xtick style={color=black},
y grid style={white!69.01960784313725!black},
ylabel={\Large Error rate},
ymajorgrids,
ymin=0, ymax=1.05,
ytick style={color=black},
width=7cm,height=8cm,
]
\addplot [very thick, dashed, blue, mark=o, mark size=4, mark options={solid}]
table {%
80 0.09085
100 0.1095
120 0.09765
140 0.06685
160 0.0475
};
\addplot [very thick, red, mark=square, mark size=4, mark options={solid}]
table {%
80 0.34215
100 0.3255
120 0.22915
140 0.19415
160 0.1375
};
\addplot [thick, color0, mark=diamond, mark size=3, mark options={solid}]
table {%
80 0.9533
100 0.98
120 0.97
140 0.98
160 0.9733
};
\addplot [thick, green!50!black, mark=x, mark size=4, mark options={solid}]
table {%
80 0.7208
100 0.74
120 0.7655
140 0.77835
160 0.782
};
\addplot [thick, black, mark=otimes, mark size=4, mark options={solid}]
table {%
80 0.544
100 0.58315
120 0.63765
140 0.69385
160 0.68815
};
\addplot [thick, orange, mark=triangle, mark size=4, mark options={solid}]
table {%
80 0.4560
100 0.5220
120 0.5320
140 0.5760
160 0.6080
};
\addplot [thick, magenta, mark=oplus, mark size=4, mark options={solid}]
table {%
80 0.9600
100 0.9820
120 0.9960
140 1.0000
160 1.0000
};
\end{axis}

\end{tikzpicture}

%% file: Evsn_wlp_ba.tex
\begin{tikzpicture}

\definecolor{color0}{rgb}{0,0.75,0.75}

\begin{axis}[
legend cell align={left},
legend columns=1,
legend style={at={(0.5,1)}, anchor=north, draw=white!80.0!black},
tick align=outside,
tick pos=both,
title={},
x grid style={white!69.01960784313725!black},
xlabel={\Large Graph size $n$},
xmajorgrids,
xmin=80, xmax=160,
xtick style={color=black},
y grid style={white!69.01960784313725!black},
ylabel={},
ymajorgrids,
ymin=0.125, ymax=0.65,
ytick style={color=black},
width=7cm,height=8cm,
]
\addplot [very thick, dashed, blue, mark=o, mark size=4, mark options={solid}]
table {%
80 0.139625
100 0.1722
120 0.2064
140 0.228425
160 0.2553125
};
\addplot [very thick, red, mark=square, mark size=4, mark options={solid}]
table {%
80 0.271125
100 0.2761
120 0.28148333
140 0.28191429
160 0.295
};
\addplot [thick, color0, mark=diamond, mark size=3, mark options={solid}]
table {%
80 0.634875
100 0.6113
120 0.58999167
140 0.56698929
160 0.566375
};
\addplot [thick, green!50!black, mark=x, mark size=4, mark options={solid}]
table {%
80 0.495625
100 0.4587
120 0.43015833
140 0.46195357
160 0.466625
};
\addplot [thick, black, mark=otimes, mark size=4, mark options={solid}]
table {%
80 0.3115
100 0.3434
120 0.37265
140 0.39162857
160 0.4100625
};
\addplot [thick, orange, mark=triangle, mark size=4, mark options={solid}]
table {%
80 0.2600
100 0.3212
120 0.3620
140 0.3826
160 0.4045
};
\addplot [thick, magenta, mark=oplus, mark size=4, mark options={solid}]
table {%
80 0.4945
100 0.5076
120 0.4767
140 0.4409
160 0.4370
};
\end{axis}

\end{tikzpicture}

%% file: Evst_time.tex
\begin{tikzpicture}

\begin{axis}[
legend cell align={left},
legend style={fill opacity=0.8, draw opacity=1, text opacity=1, at={(0.03,0.97)}, anchor=north west, draw=white!80!black},
log basis x={10},
tick align=outside,
tick pos=left,
title={},
x grid style={white!69.0196078431373!black},
xlabel={\Large Time (sec.)},
xmajorgrids,
xmin= 0.5, xmax=1900,
xmode=log,
xtick style={color=black},
y grid style={white!69.0196078431373!black},
ylabel={\Large Error rate },
ymajorgrids,
ymin=0.1, ymax=1.0,
ytick style={color=black},
width=7cm,height=7.5cm,
]
\addplot [very thick, green!50!black]
table {%
0.5 0.58
1800.50298 0.58
};
\addlegendentry{PCA \eqref{eq:pca}}
\addplot [very thick, red, mark=square, mark size=3, mark options={solid}]
table {%
80.19467916 0.48
127.20244 0.95
336.2342041 0.93
482.8130669 0.9
716.8419096 0.83
830.2459729 0.55
892.8944599 0.43
970.8837276 0.34
1019.505137 0.25
1128.710328 0.22
1184.372884 0.2
};
\addlegendentry{CVXPY}
\addplot [very thick, blue, mark=o, mark size=3, mark options={solid}]
table {%
0.548243737 0.4
1.658932734 0.39
2.766904688 0.42
8.308705091 0.37
13.28876574 0.33
19.47396049 0.3
22.28148904 0.27
26.79172082 0.25
39.8650497 0.23
54.80595577 0.2
65.91902778 0.18
83.56808281 0.16
};
\addlegendentry{Algorithm \ref{alg:nmf}}
\end{axis}

\end{tikzpicture}

%% file: Evst_iterv2.tex
\begin{tikzpicture}

\begin{axis}[
legend cell align={left},
legend style={fill opacity=0.8, draw opacity=1, text opacity=1, at={(0.03,0.97)}, anchor=north west, draw=white!80!black},
log basis x={10},
tick align=outside,
tick pos=left,
title={},
x grid style={white!69.0196078431373!black},
xlabel={\Large Iteration number},
xmajorgrids,
xmin= 0.5, xmax=1900,
xmode=log,
xtick style={color=black},
y grid style={white!69.0196078431373!black},
ymajorgrids,
ymin=0.1, ymax=1.0,
ytick style={color=black},
width=7cm,height=7.5cm,
]
\addplot [very thick, green!50!black]
table {%
0.5 0.58
1800.50298 0.58
};
\addplot [very thick, red, mark=square, mark size=3, mark options={solid}]
table {%
1 0.48
2 0.95
6 0.93
9 0.9
14 0.83
17 0.55
19 0.43
21 0.34
23 0.25
26 0.22
28 0.2
};
\addplot [very thick, blue, mark=o, mark size=3, mark options={solid}]
table {%
1	0.4
21	0.39
41	0.42
 141	0.37
 231	0.33
 341	0.3
 391	0.27
 471	0.25
 701	0.23
 971	0.2
 1171	0.18
1491	0.16
};
\end{axis}

\end{tikzpicture}

%% file: Eigenvalue_stock.tex
\begin{tikzpicture}

\definecolor{color0}{rgb}{0.12156862745098,0.466666666666667,0.705882352941177}

\begin{axis}[
tick align=outside,
tick pos=left,
x grid style={white!69.0196078431373!black},
xlabel={\large Eigenvalue index},
xmajorgrids,
ymajorgrids,
xtick style={color=black},
y grid style={white!69.0196078431373!black},
ylabel={\large Eigenvalue},
ytick style={color=black},
width=6cm,height=5cm,
ymode = log,
]
\addplot [thick, color0]
table {%
0 0.00680757271430402
1 0.00130315560195692
2 0.000917178309297768
3 0.000500830402501641
4 0.00047602253297549
5 0.000433794545342334
6 0.000385991541046359
7 0.000352516811787227
8 0.000320757926477095
9 0.000294370016768784
10 0.000274702713363628
11 0.000264656350706116
12 0.000240615733231989
13 0.000231685995786083
14 0.000208146009230258
15 0.000194651941488949
16 0.000184710464899007
17 0.000182413108469846
18 0.00017134098617464
19 0.000160126066935949
20 0.000148787135785598
21 0.000137696657759868
22 0.000130326225916226
23 0.000125055386309892
24 0.000123339026694589
25 0.000119466078284743
26 0.000116374131295011
27 0.000107717100933476
28 0.000106169264619054
29 0.000103985810498654
30 0.000103398464337141
31 9.63459013124234e-05
32 9.18963425540995e-05
33 9.01791633476153e-05
34 8.72267473234263e-05
35 8.84381922228471e-05
36 8.45403140691479e-05
37 8.159347892671e-05
38 7.80229120445046e-05
39 7.09941857198956e-05
40 7.30088091493881e-05
41 7.21202232667732e-05
42 6.5678584338714e-05
43 6.76956320233338e-05
44 6.37715912631095e-05
45 6.22450590756816e-05
46 5.94185352359139e-05
47 5.85294458082009e-05
48 5.66193727863112e-05
49 5.51079504027352e-05
50 1.26431341869523e-06
51 5.19927788818214e-05
52 5.04529474965967e-05
53 4.93368100638423e-05
54 4.77669717732954e-05
55 4.68703780380574e-05
56 4.49697725052205e-05
57 4.88585073280299e-06
58 6.08013220285481e-06
59 6.37058350632366e-06
60 4.26584647045587e-05
61 4.19433466347789e-05
62 4.11504691887756e-05
63 3.92537983797414e-05
64 7.22352987441495e-06
65 7.64852083088318e-06
66 3.76969400984767e-05
67 8.71145370502597e-06
68 9.28032469580037e-06
69 9.85020800335307e-06
70 1.01247843715806e-05
71 3.41694727994116e-05
72 3.27819528871916e-05
73 3.14882295618749e-05
74 3.04129008007113e-05
75 1.11375202868109e-05
76 2.94527424672383e-05
77 1.20009895271458e-05
78 1.24467974650577e-05
79 2.75085908031687e-05
80 2.67439241915266e-05
81 2.5952644672688e-05
82 1.32610978114655e-05
83 1.54879180797291e-05
84 1.58127461166975e-05
85 1.34529836899072e-05
86 2.27408034432206e-05
87 3.59134683306125e-05
88 3.37791347748456e-05
89 2.41922037189944e-05
90 1.7254367732774e-05
91 1.76764149725072e-05
92 2.33430399645001e-05
93 2.05591550579659e-05
94 1.99744177127801e-05
95 1.93561982593244e-05
96 1.85878973627244e-05
97 1.26846039817475e-05
98 1.90666798781546e-05
};
\addplot [thick, black, dashed]
table { 
20 1e-06
20 1e-02
};
\end{axis}

\node (test) at (1.65,3) {k=20};

\end{tikzpicture}

%% file: Eigenvalue_vote.tex
\begin{tikzpicture}

\definecolor{color0}{rgb}{0.12156862745098,0.466666666666667,0.705882352941177}

\begin{axis}[
tick align=outside,
tick pos=left,
x grid style={white!69.0196078431373!black},
xlabel={\large Eigenvalue index},
xmin=-2.45, xmax=51.45,
xtick style={color=black},
y grid style={white!69.0196078431373!black},
ytick style={color=black},
xmajorgrids,
ymajorgrids,
width=6cm,height=5cm,
ymode = log,
]
\addplot [thick, color0]
table {%
0 14.5765792951419
1 8.76135567187959
2 0.858219077521218
3 0.716109093839221
4 0.521762758392826
5 0.428668085025198
6 0.414492075070271
7 0.390805782806404
8 0.331450763320535
9 0.302138250874947
10 0.288842612030804
11 0.274256864644172
12 0.258472838688725
13 0.244245132237523
14 0.23426940336767
15 0.225627208950483
16 0.218601002388333
17 0.209583447933914
18 0.203740144591177
19 0.192023251973613
20 0.0437365621363346
21 0.17987068019301
22 0.17258219369174
23 0.168913857454521
24 0.0572925744653427
25 0.0604827254780003
26 0.0617151648135257
27 0.0642776482902188
28 0.0675178841000499
29 0.159302927813663
30 0.158219199443137
31 0.0746741752408997
32 0.0763811008612555
33 0.0799904745592143
34 0.150196054562358
35 0.140769202934632
36 0.141327500266262
37 0.135038713364721
38 0.12941351757224
39 0.0883646288962311
40 0.124678682144634
41 0.120099783374972
42 0.116378816730007
43 0.0942164268133046
44 0.112067945515109
45 0.0976318019895984
46 0.108401408450974
47 0.105051997947249
48 0.10101764003022
49 0.102458178126365
};
\addplot [thick, black, dashed]
table { 
10 1e-01
10 1e+01
};
\end{axis}

\node (test) at (1.65,2.75) {k=10};

\end{tikzpicture}

%% file: consist.tex
\begin{tikzpicture}

\definecolor{color0}{rgb}{0,0.75,0.75}

\begin{axis}[
legend cell align={left},
legend style={fill opacity=0.8, draw opacity=1, text opacity=1, draw=white!80.0!black},
tick align=outside,
tick pos=left,
x grid style={white!69.01960784313725!black},
xlabel={\large Amount of available data},
xmajorgrids,
xmin=-0.2, xmax=4.2,
xtick style={color=black},
xtick={0,1,2,3,4},
xticklabels={$0.1m$,$0.3m$,$0.5m$, $0.7m$, $0.9m$},
y grid style={white!69.01960784313725!black},
ylabel={$||{\bm \mu}_{\sf full} - \widehat{\bm \mu}||_1$},
ymajorgrids,
ymin=-0.0254746808878043, ymax=1.4,
ytick style={color=black},
width=9cm,height=5cm,
]
\addplot [very thick, red, mark=o, mark size=4, mark options={solid}]
table {%
0 0.71506
1 0.65148
2 0.46228
3 0.28536
4 0.0384
};
\addlegendentry{{\sf Senate}}

\addplot [very thick, blue, mark=square, mark size=4, mark options={solid}]
table {%
0 1.29628
1 0.91206
2 0.48524
3 0.219292727272727
4 0.08942
};
\addlegendentry{{\sf Stock}}

\end{axis}

\end{tikzpicture}